\documentclass[]{article}       
\usepackage[onecolumn]{apjgalley}
\usepackage{psfig}
\journalid{VOL}{JOURNAL DATE 1997}
\articleid{START PAGE}{END PAGE}
\submitted{Astrophysical Journal, to appear in January 10, 2000 issue,
 Vol. 528}
\font\fiverm=cmr5             \font\sevenrm=cmr7

          \font\sixrm=cmr6
               
\def\pr{Phys. Rev.}                             
\def\jetp{Sov. Phys. JETP}                      
\def\rmp{Rev. Mod. Phys.}                       
\def\ssr{Space Sci. Rev.}                       
\def\teq#1{$\, #1\,$}                           
\def\today{\ifcase\month\or
  January\or February\or March\or April\or May\or June\or
  July\or August\or September\or October\or November\or
  December\fi
  \space\number\day, \number\year}
{\catcode`\@=11
\gdef\SchlangeUnter#1#2{\lower2pt\vbox{\baselineskip 0pt\lineskip0pt
\ialign{$\m@th#1\hfil##\hfil$\crcr#2\crcr\sim\crcr}}}}
\def\gtrsim{\mathrel{\mathpalette\SchlangeUnter>}}
\def\lesssim{\mathrel{\mathpalette\SchlangeUnter<}}

\def\fsc{\alpha_{\hbox{\sevenrm f}}}

\def\dover#1#2{\hbox{${{\displaystyle#1 \vphantom{(} }\over{
   \displaystyle #2 \vphantom{(} }}$}}

\begin{document}
%
%
\newcommand{\vol}[2]{$\,$\rm #1\rm , #2.}
\newcommand{\figureout}[2]{ \figcaption[#1]{#2} }       
%
\newcommand{\tableout}[4]{\vskip 0.3truecm \centerline{\rm TABLE #1\rm}
   \vskip 0.2truecm\centerline{\rm #2\rm}   
   \vskip -0.3truecm  \begin{displaymath} #3 \end{displaymath}
   \noindent \rm #4\rm\vskip 0.1truecm } 
%
%
%
\title{INVERSE BREMSSTRAHLUNG IN SHOCKED ASTROPHYSICAL PLASMAS}
   \author{Matthew G. Baring\altaffilmark{1} and Frank C. Jones}
   \affil{Laboratory for High Energy Astrophysics, Code 661, \\
      NASA Goddard Space Flight Center, Greenbelt, MD 20771, U.S.A.\\
      \it baring@lheavx.gsfc.nasa.gov, frank.c.jones@gsfc.nasa.gov\rm}

   \and

   \author{Donald C. Ellison}
   \affil{Department of Physics, North Carolina State University,\\
          Box 8202, Raleigh NC 27695, U.S.A.\\
          \it don\_ellison@ncsu.edu\rm}

   \altaffiltext{1}{Universities Space Research Association}
   \authoraddr{Laboratory for High Energy Astrophysics, Code 661,
      NASA Goddard Space Flight Center, Greenbelt, MD 20771, U.S.A.}
%
%
%
%
\begin{abstract}
There has recently been interest in the role of inverse bremsstrahlung,
the emission of photons by fast suprathermal ions in collisions with
ambient electrons possessing relatively low velocities, in tenuous
plasmas in various astrophysical contexts.  This follows a long hiatus
in the application of suprathermal ion bremsstrahlung to astrophysical
models since the early 1970s.  The potential importance of inverse
bremsstrahlung relative to normal bremsstrahlung, i.e. where ions are
at rest, hinges upon the underlying velocity distributions of the
interacting species.  In this paper, we identify the conditions under
which the inverse bremsstrahlung emissivity is significant relative to
that for normal bremsstrahlung in shocked astrophysical plasmas.  We
determine that, since both observational and theoretical evidence
favors electron temperatures almost comparable to, and certainly not
very deficient relative to proton temperatures in shocked plasmas,
these environments generally render inverse bremsstrahlung at best a
minor contributor to the overall emission.  Hence inverse
bremsstrahlung can be safely neglected in most models invoking shock
acceleration in discrete sources such as supernova remnants.  However,
on scales \teq{\gtrsim 100}pc distant from these sources, Coulomb
collisional losses can deplete the cosmic ray electrons, rendering
inverse bremsstrahlung, and perhaps bremsstrahlung from knock-on
electrons, possibly detectable.

\end{abstract}
%
%
\section{INTRODUCTION}
 \label{sec:intro}

The process of inverse bremsstrahlung has received attention from time
to time over the last three decades, in various astrophysical
settings.  Inverse bremsstrahlung is defined here to be the emission of
a single photon when a high speed ion collides with an electron that is
effectively at rest, and has often been referred to as suprathermal
proton bremsstrahlung (e.g. Boldt \& Serlemitsos 1969; Brown 1970;
Jones 1971; Haug 1972).   Being a kinematic inverse (an analogy can be
drawn with inverse Compton scattering) arising purely via Lorentz
transformations between reference frames (i.e. it is still the electron
that radiates), it is not to be confused with the true quantum
electrodynamical inverse of conventional bremsstrahlung, namely the
absorption of photons in 3-body collisions with electrons and ions, a
process that is highly improbable in tenuous (i.e. optically thin)
astrophysical plasmas.  

The first application of suprathermal proton bremsstrahlung in
astrophysical models dates to the work of Hayakawa \& Matsuoka (1964),
which considered such radiation in collisions between cosmic rays and
ambient electrons in the intergalactic medium as a source of the cosmic
X-ray background (XRB) at energies \teq{\lesssim 10}keV.  This concept
was developed in the papers by Hayakawa (1969), Boldt \& Serlemitsos
(1969), and Brown (1970), with Boldt \& Serlemitsos proposing that
suprathermal proton bremsstrahlung could account for the flat spectral
index of the XRB below around 15 keV, while noting that an unusually
high column density of ambient electrons would be required to match the
observed flux.  The interest in the relevance of inverse bremsstrahlung
to the cosmic XRB has since diminished, and more recent opinion favours
the accumulation of unresolved discrete sources as the dominant
contributions (see Fabian \& Barcons 1992, for a comprehensive
review).  Although thermal bremsstrahlung at around 40 keV can fit the
2--100 keV spectrum fairly well (Marshall, et al. 1980; see also Boldt
1987, for a review), recent critical evidence limits the contribution
of truly diffuse emission from the intergalactic medium to less than
0.01\% of the observed flux.  This bound is derived from upper
limits to the mean cosmological electron density obtained from the
measurements of COBE to Compton scattering-induced distortion of the
cosmic microwave background (Wright, et al. 1994).

Renewed interest in inverse bremsstrahlung has arisen in the last
year.  Tatischeff, Ramaty \& Kozlovsky (1998) and Dogiel et al. (1998)
have discussed expectations for suprathermal proton bremsstrahlung in
the X-ray band, assuming a significant enhancement of the density of
low energy cosmic ray ions in the Orion region.  Inferences of such an
abundance of ions were drawn from the reported (e.g. Bloemen et al. 1994),
but recently retracted (Bloemen et al. 1999), detection with
COMPTEL/CGRO of emission lines from the Orion molecular cloud complex;
these were attributed to Carbon and Oxygen nuclear line emission in
collisions with ambient ions.  In a different context, Valinia and
Marshall (1998) conjectured that inverse bremsstrahlung from cosmic rays
could be responsible for much of the diffuse X-ray emission seen in the
galactic ridge.  Pohl (1998) has responded to this proposition by
arguing that it is difficult to produce the diffuse component in the
ridge via inverse bremsstrahlung without violating observational limits
to nuclear excitation line and pion decay continuum emission.
Furthermore, Tatischeff, Ramaty \& Valinia (1999) argue that by
normalizing the low energy cosmic ray ion populations using the
galactic Be production rate, the contribution of inverse bremsstrahlung
to the thin disk component of unresolved/diffuse 10-60 keV emission
detected by RXTE is at most a few percent.

The focus of the presentation here is on collisionless shocked
astrophysical plasmas, which distinguish themselves from the diffuse
emission scenarios addressed by Valinia and Marshall (1998), Pohl
(1998) and Tatischeff, Ramaty \& Valinia (1999) by the insignificance
of Coulomb collisions compared to interactions between the magnetic
field and charged particles.  The principal examples in mind here are
young supernova remnants, i.e.  those perhaps mature enough to be in
the Sedov epoch, but well before the radiative phase.  The
bremsstrahlung emission properties of shocked environments are dictated
by the dissipation between ions and electrons in the shock layer.  We
show in Section~\ref{sec:ibveb} that only when electron and ion thermal
speeds are comparable (i.e., when the temperature ratio is on the order
of, or less than the mass ratio, \teq{T_e / T_p \lesssim m_e / m_p})
can inverse bremsstrahlung contribute significantly relative to normal
electron bremsstrahlung in the optical, X-ray and gamma-ray wavebands.
We argue in Subsection~\ref{sec:expect} that such conditions are not
easily realized in the majority of {\it shocked} astrophysical
environments, thereby rendering suprathermal proton bremsstrahlung
unimportant for most discrete sources.  By extension, the importance of
inverse bremsstrahlung as a continuum emission mechanism in extended
regions is contingent upon there being a relative paucity of primary
cosmic ray electrons in the interstellar medium on appropriate length
scales (\teq{\gtrsim 100}pc).  This point, and the role of knock-on
electron bremsstrahlung, are discussed in Section~\ref{sec:discuss}.

\section{INVERSE BREMSSTRAHLUNG VERSUS ELECTRON BREMSSTRAHLUNG}
 \label{sec:ibveb}

The key question of interest to astrophysicists that we address here is
when the inverse bremsstrahlung process is significant relative to the
normal bremsstrahlung mechanism in shocked astrophysical plasmas.
Clearly, the relative numbers of projectile and target particles will
be central to answering this question.  Here we aim to determine how
such numbers and the associated emissivities for the two processes
depend on standard plasma parameters.  The emissivities, or photon
production rates for either process, can be written in the form
\begin{equation}
   \dover{dn_{\gamma}(\omega )}{dt}\; =\; n_t
   \int_{0}^{\infty} dp\, n_{\hbox{\fiverm CR}}(p)\,
   c\beta\, \dover{d\sigma}{d\omega}\quad ,
 \label{eq:prodrate}
\end{equation}
for shock-accelerated cosmic ray (projectile) particles with momentum
distribution \teq{n_{\hbox{\fiverm CR}}(p)} colliding with targets of
density \teq{n_t} that are effectively at rest in the observer's
frame.  Here \teq{d\sigma /d\omega} is the cross-section, differential
in the photon energy, where we adopt the convention of using
dimensionless units throughout this paper: photon energies
\teq{E_{\gamma}} are expressed in units of the electron rest mass
\teq{m_ec^2}, i.e., via \teq{\omega=E_{\gamma}/(m_ec^2)}.  Hence
\teq{d\sigma /d\omega} has c.g.s. units of cm$^2$.  Also, \teq{c\beta}
is the speed of the projectile of momentum \teq{p}.

The distribution functions of accelerated particles in shock environs
need to be described numerically, in general, as will be evident from
the Monte Carlo-generated distributions illustrated later in the
paper.  However, we can provide a good estimate of the relative
importance of inverse bremsstrahlung and classical bremsstrahlung using
simple but representative analytic forms for the particle distribution
functions.  For these purposes, let us approximate the cosmic ray
(shock-accelerated) proton and electron distributions in a discrete
source by power-laws in momentum, broken at thermal energies.  Defining
\teq{c\beta_{\hbox{\fiverm T},p}} and \teq{c\beta_{\hbox{\fiverm T},e}}
to be the proton and electron thermal speeds, respectively, their
momentum distributions can be cast in the approximate form
\begin{equation}
   n_s(p) \; =\; \dover{n_s}{p_{\hbox{\fiverm T},s}}\;
   \;\dover{3(\Gamma_s-1)}{\Gamma_s-1+3\epsilon_s}
   \cases{  \biggl( \dover{p}{p_{\hbox{\fiverm T},s}} \biggr)^2\; ,
              & $p\,\leq\, p_{\hbox{\fiverm T},s}$ \cr
            \epsilon_s\;
            \biggl( \dover{p}{p_{\hbox{\fiverm T},s}} \biggr)^{-\Gamma_s}\; ,
              & $p\, > \, p_{\hbox{\fiverm T},s}$ \cr }
 \label{eq:crdists}
\end{equation}
where \teq{p_{\hbox{\fiverm T},s}=m_sc\beta_{\hbox{\fiverm T},s}} is
the thermal (non-relativistic) momentum for species \teq{s}
(\teq{=e,p}).  Here, \teq{n_s} represents the total density and
\teq{\Gamma_s} is the power-law index of a given {\it projectile}
species.  This discontinuous ``schematic'' form is chosen
to mimic the {\it slopes} of the low momentum portion of a
Maxwell-Boltzmann distribution and the non-thermal suprathermal portion
of the cosmic ray distribution, and hence does not accurately describe
the spectral structure near the thermal peak.  Consideration of such
spectral details, which depend on the nature of dissipation in the
shock layer, would introduce only modest corrections to the rates
derived below, and are immaterial to the qualitative conclusions of
this paper.  The factor \teq{\epsilon_s} is introduced to account for
the fact that the non-thermal tail does not extrapolate directly from
the thermal peak in shock acceleration-generated populations, and its
values \teq{\lesssim 1} define the efficiency of the acceleration
mechanism, which can be a strong function of the field obliquity (e.g.
Baring, Ellison \& Jones 1993) and the strength of particle scattering
(e.g. Ellison, Baring \& Jones 1996) in the shock environs.  For ions,
acceleration is generally quite efficient, implying \teq{\epsilon_p\sim
0.1}.  For electrons, the situation is less clear, pertaining to the
well-known electron injection problem in non-relativistic shocks.
However, as will be mentioned in Subsection~\ref{sec:expect} below, the
observed electron-to-proton cosmic ray abundance ratio provides a
canonical lower bound to the injection efficiency of electrons, and
generally implies that \teq{\epsilon_e} cannot be dramatically less than
\teq{\epsilon_p}.  Finally, note that the target density of cool
particles for the two processes in question will be written as
\teq{n_t}, and will represent a subset of either of the accelerated
populations that are formed from the shock heating of the interstellar
medium.

\subsection{The Bremsstrahlung Ratio in a Nutshell}
 \label{sec:bremsrat}

The ratio \teq{{\cal R}_{\rm O-X}} of inverse bremsstrahlung to
bremsstrahlung emission in the optical to X-ray and soft gamma-ray
bands can be quickly written down without the encumbrance of the
mathematical complexity of the differential cross-sections involved.
Here we enunciate this, the principal result of the paper in a simple
and enlightening manner, so that the detailed expositions of the next
two subsections can be bypassed by the reader, if desired.  At energies
below around a few hundred keV, the waveband of interest to the
discussions of Boldt \& Serlemitsos (1969), Hayakawa (1969), Brown
(1970), Tatischeff, Ramaty \& Kozlovsky (1998), Dogiel et al. (1998)
and Valinia and Marshall (1998), the non-relativistic differential
cross-section in Eq.~(\ref{eq:csect_nr}) below is operable, and applies
to both bremsstrahlung and inverse bremsstrahlung.  This cross-section
is a function only of the speed of the ballistic particle involved, for
a given photon energy.  Hence, any disparity in the emissivities for
the two processes can only be due to differences in the numbers of
ballistic protons and electrons at a given speed, a property that
does not extend to relativistic domains.  Since bremsstrahlung
X-rays in supernova remnant shocks are generally produced by slightly
superthermal electrons, this number is just
\teq{n_e\epsilon_e/\beta_{\hbox{\fiverm T},e}}, reflecting the
efficiency of injection into the shock acceleration process.  For
protons, whose thermal speeds are much lower, only a small fraction can
participate in inverse bremsstrahlung collisions with shocked
electrons, namely those with speeds exceeding the electron thermal
speed \teq{\beta_{\hbox{\fiverm T},e}}.  Using the suprathermal portion
of Eq.~(\ref{eq:crdists}), this constitutes a ``density''
\teq{n_p\epsilon_p\, \beta_{\hbox{\fiverm T},p}^{\Gamma_p-1}
\beta_{\hbox{\fiverm T},e}^{-\Gamma_p}}, so that it immediately follows
that the ratio of the two processes is of the order of
\begin{equation}
   {\cal R}_{\rm O-X} \; =\; \dover{\epsilon_p}{\epsilon_e}\;
  \biggl( \dover{\beta_{\hbox{\fiverm T},p}}{
                 \beta_{\hbox{\fiverm T},e}} \biggr)^{\Gamma_p-1}\;\; ,
 \label{eq:ratio_O-X}
\end{equation}
for \teq{n_e\sim n_p}, as required by charge neutrality.  Therein lies
the principal result, borne out in the derivations of the next
two subsections, which are expounded because of their usefulness in
astrophysical problems.  Clearly, the ratio \teq{{\cal R}_{\rm O-X}}
depends principally on the degree of temperature equilibration, or
otherwise, in the shocked plasma.

\subsection{Radiation from Non-relativistic Particles}
 \label{sec:ibnr}

A more-detailed estimate of the relative importance of inverse
bremsstrahlung and classical bremsstrahlung at optical to X-ray
energies can be obtained by considering only non-relativistic
accelerated ions, i.e. those with \teq{\beta\ll 1}.  In this regime,
the differential cross-section \teq{d\sigma /d\omega} for either
bremsstrahlung or its inverse can be obtained from textbooks such as
Jauch and Rohrlich (1980), and is the non-relativistic specialization
of the Bethe-Heitler formula:
\begin{equation}
   \dover{d\sigma}{d\omega}\biggl\vert_{\hbox{\sixrm BH}}\; =\;
   \dover{16}{3}\, Z^2\,
   \dover{\fsc\, r_0^2}{\omega} \,\dover{1}{\beta^2}\,\log_e
   \dover{\beta +\sqrt{\beta^2-2\omega}}{\beta -\sqrt{\beta^2-2\omega}}
   \quad , \quad 2\pi\fsc Z\lesssim \beta\ll 1\;\; ,
 \label{eq:csect_nr}
\end{equation}
where \teq{\fsc =e^2/(\hbar c)} is the fine structure constant,
\teq{+Ze} is the nuclear charge, and \teq{r_0=e^2/(m_ec^2)} is the
classical electron radius.  Here \teq{\beta} is the relative speed (in
units of \teq{c}) between the projectile and target, namely the
electron speed for classical bremsstrahlung, and the proton (or ion)
speed for the inverse process.  The applicability of this formula,
which is integrated over final electron and photon angles, for both
bremsstrahlung and its inverse follows from the Lorentz invariance of
the total cross-section \teq{\sigma} and the effective invariance of
photon angles and energies in non-relativistic transformations between
the proton and electron rest frames.  With this interpretation,
\teq{c\sqrt{\beta^2-2\omega}} represents the final electron speed in
the proton rest frame for either process.  Note that
Eq.~(\ref{eq:csect_nr}) is derived in the Born or plane-wave
approximation, and is strictly not applicable to regimes where
\teq{\beta\lesssim 2\pi\fsc Z} or \teq{\omega\ll\beta^2\ll 1} where
Coulomb perturbations to the electron wave-functions become important
(see the end of this section).  Appropriate results for low frequencies
and classical regimes can be found in Gould (1970), and references
therein, and amount to modifications of the argument of the logarithm.

Using Equation~(\ref{eq:prodrate}), it is straightforward to write down
production rates of photons for the two processes in the limit of
non-relativistic projectile speeds.  For regular bremsstrahlung, the
vast majority of shock-accelerated protons are effectively cold due to their
generally low thermal speed (except for unusually
hot proton components in tenuous plasmas), so that the proton density
\teq{n_p} represents the target density \teq{n_t}.  Hence the photon
production rate is
\begin{equation}
   \dover{dn_{\gamma}(\omega )}{dt}\biggl\vert_{\rm brems}\; \approx\;
   \dover{16}{3}\, \dover{n_pn_e}{\beta_{\hbox{\fiverm T},e}} \, Z^2\,
   \dover{\fsc\, r_0^2c}{\omega}\,
   \dover{3(\Gamma_e-1)}{\Gamma_e-1+3\epsilon_e}\,
   \Bigl\{ f_1(\omega ,\, \beta_{\hbox{\fiverm T},e}, \, \Gamma_e) +
   \epsilon_e\, f_2(\omega ,\, \beta_{\hbox{\fiverm T},e}, \, \Gamma_e) \Bigr\}
   \quad ,
 \label{eq:nrbrems}
\end{equation}
where \teq{\Gamma_e} is the electron power-law spectral index, and
\begin{eqnarray}
   f_1(\omega ,\, \beta_{\hbox{\fiverm T}}, \, \Gamma) & = &
   \dover{1}{\beta_{\hbox{\fiverm T}}^2}
   \int_{\sqrt{2\omega}}^{\beta_c} d\beta\, \beta\,
   \log_e \dover{\beta +\sqrt{\beta^2-2\omega}}{\beta -\sqrt{\beta^2-2\omega}}
      \nonumber\\
   & = &
   \dover{1}{\beta_{\hbox{\fiverm T}}^2}\,\Biggl\{  (\beta_c^2-\omega)
    \log_e \dover{\beta_c +\sqrt{\beta_c^2-2\omega}}{\sqrt{2\omega}}
    - \dover{\beta_c}{2} \, \sqrt{\beta_c^2-2\omega} \Biggr\}\;\; ,
\label{eq:fdef}\\
   f_2(\omega ,\, \beta_{\hbox{\fiverm T}}, \, \Gamma) & = &
   \beta_{\hbox{\fiverm T}}^{\Gamma} \int^{1}_{\beta_c}
     \dover{d\beta}{\beta^{1+\Gamma}}\,
   \log_e \dover{\beta +\sqrt{\beta^2-2\omega}}{\beta -\sqrt{\beta^2-2\omega}}
   \;\; ,\nonumber
\end{eqnarray}
for
\begin{equation}
   \beta_c\; =\; \max\Big\{\,
   \beta_{\hbox{\fiverm T}}, \; \sqrt{2\omega}\, \Bigr\}\; .
 \label{eq:betacdef}
\end{equation}
Here we use the argument \teq{\Gamma} to represent either
\teq{\Gamma_e} or \teq{\Gamma_p}, as the case may be, in the work
below.  The first integral in Equation~(\ref{eq:fdef}) specializes to
\teq{[\log_e(2 \beta_{\hbox{\fiverm T}}^2/\omega)-1]/2} in the limit of
\teq{\beta_{\hbox{\fiverm T}}^2\gg\omega}.  The second integral in
Equation~(\ref{eq:fdef}) is generally expressible in terms of
hypergeometric functions, though such a manipulation is not
enlightening.  Specialization to the \teq{\beta_{\hbox{\fiverm
T}}^2\gg\omega} limit yields tractable integrals and the result
\teq{[\log_e(2 \beta_{\hbox{\fiverm T}}^2/\omega)+2/\Gamma]/\Gamma}.
However, remembering that modest corrections to the cross-section in
Equation~(\ref{eq:csect_nr}) are required (Gould 1970) in this low
frequency limit, we restrict use of this specialization to the range
\teq{\beta_{\hbox{\fiverm T}}^2\gtrsim\omega}.  The other interesting
limit is when \teq{\beta_{\hbox{\fiverm T}}^2\ll\omega\ll 1}, for which
only the second integral contributes; the result in this case is
\teq{f_2(\omega ,\, \beta_{\hbox{\fiverm T}}, \, \Gamma)\approx
2^{\Gamma -2}\, B(\Gamma,\,\Gamma )}, where \teq{B(x,x)} is the beta
function.  The substitutions \teq{\beta = \sqrt{2\omega}\,\cosh\theta},
and identity 3.512.1 in Gradshteyn and Ryzhik (1980) facilitate these
developments.  The limiting cases can then be summarized as
\begin{equation}
   \dover{dn_{\gamma}(\omega )}{dt}\biggl\vert_{\rm brems}\; \approx\;
   16\, \dover{\Lambda}{\beta_{\hbox{\fiverm T},e}} \,
   \dover{\Gamma_e-1}{\Gamma_e-1+3\epsilon_e}\,
   \cases{ \dover{1}{\omega} \,
          \biggl\{ \Bigl( \dover{1}{2} + \dover{\epsilon_e}{\Gamma_e}\Bigr)
            \log_e \dover{2 \beta_{\hbox{\fiverm T},e}^2}{\omega}
           - \dover{1}{2} + \dover{2\epsilon_e}{\Gamma_e^2} \biggr\}\, , &
               $\omega \lesssim \beta_{\hbox{\fiverm T},e}^2 \ll 1$, \cr
           \epsilon_e\; \dover{2^{\Gamma_e/2-1}}{\omega^{1+\Gamma_e/2}}\;
           \beta_{\hbox{\fiverm T},e}^{\Gamma_e}\,  \vphantom{\Biggl(}
           \dover{\Gamma^2(\Gamma_e)}{\Gamma (2\Gamma_e)}\;\; , &
               $\beta_{\hbox{\fiverm T},e}^2 \ll \omega \ll 1$,        }
 \label{eq:nrbrems_lim}
\end{equation}
where \teq{\Gamma(x)} is the Gamma function, and
\begin{equation}
   \Lambda \; =\; Z^2 n_pn_e \fsc\, r_0^2c\,
 \label{eq:Lambdadef}
\end{equation}
defines the fundamental scale for the bremsstrahlung emissivity (i.e.
at \teq{\omega \sim 1} and \teq{\beta_{\hbox{\fiverm T}}\sim 1}).  The
\teq{\omega^{-1}} behaviour at low energies reflects the convolution of
the bremsstrahlung infra-red divergence with a distribution possessing
a finite number of particles.  The \teq{\omega^{-(1+\Gamma_e/2)}}
dependence above electron thermal energies arises because the
bremsstrahlung flux spectrum traces the electron {\it energy} distribution.
Note that Eq.~(\ref{eq:nrbrems_lim}) is strictly valid only when
\teq{\Gamma_e >1}, which will always be the case for convergent electron
distributions.  A simplified derivation of the form of Eq.~(\ref{eq:nrbrems_lim}) is presented in the Appendix.

The inverse bremsstrahlung spectrum can be evaluated in a similar
manner, but noting that zero contribution arises whenever the proton
speed drops below the electron thermal speed; such a domain of phase
space corresponds to normal bremsstrahlung.  Hence, thermal protons are
never sampled unless the proton thermal speed is comparable to that of
the electrons.  It follows that the emissivity resembles the form in
Equation~(\ref{eq:nrbrems}) with just an \teq{f_2}-type term;
extracting the appropriate power-law factor, the inverse bremsstrahlung
emissivity can be written
\begin{equation}
   \dover{dn_{\gamma}(\omega )}{dt}\biggl
   \vert_{\hbox{\sevenrm inv}\atop \hbox{\sevenrm brems}} \; \approx\;
   \dover{16}{3}\, \dover{n_pn_e}{\beta_{\hbox{\fiverm T},p}} \, Z^2\,
   \dover{\fsc\, r_0^2c}{\omega}\,
   \dover{3(\Gamma_p-1)}{\Gamma_p-1+3\epsilon_p}\,
   \biggl( \dover{\beta_{\hbox{\fiverm T},p}}{\beta_{\hbox{\fiverm T},e}}
          \biggr)^{\Gamma_p}\, \epsilon_p\,
   f_2(\omega ,\, \beta_{\hbox{\fiverm T},e}, \, \Gamma_p) \quad ,
 \label{eq:nrinvbrems}
\end{equation}
where $\Gamma_p$ is the proton power-law index.
The limiting forms of the \teq{f_2} function are just as for the
bremsstrahlung case, so that we quickly arrive at the limits
\begin{equation}
   \dover{dn_{\gamma}(\omega )}{dt}\biggl
   \vert_{\hbox{\sevenrm inv}\atop \hbox{\sevenrm brems}} \; \approx\;
   16\, \dover{\Lambda}{\beta_{\hbox{\fiverm T},e}} \,
   \dover{\Gamma_p-1}{\Gamma_p-1+3\epsilon_p}\,
   \biggl( \dover{\beta_{\hbox{\fiverm T},p}}{\beta_{\hbox{\fiverm T},e}}
          \biggr)^{\Gamma_p-1}\,
   \cases{ \dover{1}{\omega}\; \dover{\epsilon_p}{\Gamma_p}\,
          \biggl\{ \log_e \dover{2 \beta_{\hbox{\fiverm T},e}^2}{\omega}
           + \dover{2}{\Gamma_p} \biggr\}\, , &
               $\omega \lesssim \beta_{\hbox{\fiverm T},e}^2 \ll 1$, \cr
           \epsilon_p\, \dover{2^{\Gamma_p/2-1}}{\omega^{1+\Gamma_p/2}}\;
           \beta_{\hbox{\fiverm T},e}^{\Gamma_p}\,  \vphantom{\Biggl(}
           \dover{\Gamma^2(\Gamma_p)}{\Gamma (2\Gamma_p)}\;\; , &
               $\beta_{\hbox{\fiverm T},e}^2 \ll \omega \ll 1$.        }
 \label{eq:nrinvbrems_lim}
\end{equation}
The two photon ranges of interest here, namely \teq{\omega \lesssim
\beta_{\hbox{\fiverm T},e}^2 \ll 1} and \teq{\beta_{\hbox{\fiverm
T},e}^2 \ll \omega \ll 1}, are the same as those for bremsstrahlung.
This is due in part to the fact that only protons with speeds above
electron thermal speeds will participate in inverse bremsstrahlung
interactions.  Note that relaxing the \teq{\beta_p\gtrsim
\beta_{\hbox{\fiverm T},e}} requirement to bounds on \teq{\beta_p}
greater than \teq{\beta_{\hbox{\fiverm T},e}} would introduce the
appropriate numerical factor in Equation~(\ref{eq:nrinvbrems_lim}).
Clearly, the factor \teq{[\epsilon_p /\epsilon_e](\beta_{\hbox{\fiverm
T},p} / \beta_{\hbox{\fiverm T},e})^{\Gamma_p-1}} roughly defines the
ratio of inverse bremsstrahlung to bremsstrahlung emissivities;
expectations for its value in shocked plasmas will form the center of
the discussion in Section~\ref{sec:expect} below.

It is salient to remark at this point that the differential
cross-section in Equation~(\ref{eq:csect_nr}) that we are using (and
also the ultra-relativistic limit in Eq.~(\ref{eq:csect_er}) below of the
Bethe-Heitler result) was obtained from QED calculations in the Born
approximation.  At non-relativistic speeds, the Coulomb potential of
the target electron or ion perturbs the projectile electron
wave-function sufficiently that the plane-wave approximation breaks
down.  This occurs when \teq{p_e\, x} drops below \teq{\hbar},
where \teq{p_e=\beta m_e c} is the momentum of the energetic
electron.  Here \teq{x} is the spatial scale appropriate to the
interaction, which is the classical electron radius \teq{r_0}.  Hence,
so-called Coulomb corrections become necessary when \teq{\beta\lesssim
\fsc =e^2/(\hbar c)}.  In such cases, the matrix element for
scattering must be determined using Coulomb wave functions.  This has
been done exactly for non-relativistic electrons colliding with nuclei
by Sommerfeld (1931), leading to the application of a simple corrective
multiplicative factor (Elwert 1939)
\begin{equation}
   {\cal C}_{\hbox{\sixrm SE}}(\beta,\,\omega ) \; =\; \dover{\beta}{\beta'}\;
   \dover{1-\exp\Bigl[ -2\pi\fsc Z/\beta \Bigr]}{
          1-\exp\Bigl[ -2\pi\fsc Z/\beta' \Bigr]}
   \quad , \quad \beta'=\sqrt{\beta^2-2\omega}\;\; ,
 \label{eq:SEcorr}
\end{equation}
known as the Sommerfeld-Elwert factor, to the Bethe-Heitler
cross-section:
\begin{equation}
   \dover{d\sigma}{d\omega}\biggl\vert_{\hbox{\sixrm SE}}\; =\;
   {\cal C}_{\hbox{\sixrm SE}}(\beta,\,\omega )\,
   \dover{d\sigma}{d\omega}\biggl\vert_{\hbox{\sixrm BH}}
 \label{eq:SEcsect}
\end{equation}
Here \teq{\beta'} is the electron's speed (in units of \teq{c}) in the
nuclear rest frame after collision, and \teq{Z} is the charge number of
the nucleus.  Note that a similar factor can be employed for
electron-electron collisions (Maxon and Corman 1967, see also Haug
1975).

Such factors become important for projectile electron speeds below
around \teq{c/10} (for proton targets).  When \teq{\beta_{\hbox{\fiverm
T},e}\ll 2\pi\fsc Z}, the Coulomb correction factor is simply
\teq{\beta/\sqrt{\beta^2-2\omega}}, and it propagates through all the
integrands of the above developments.  It is then quickly ascertained
that in the \teq{\omega \lesssim \beta_{\hbox{\fiverm T},e}^2 \ll 1}
limiting cases, the correction factor is close to unity and therefore
can be neglected (modest low frequency corrections for \teq{\omega \ll
\beta^2} are discussed in Gould 1970).  For higher photon energies, the
Coulomb corrections should be included, with the appropriate
modification to the algebraic manipulations.  While the resulting
integral is not tractable except in terms of higher order
hypergeometric functions, we determined numerically that the corrective
factor that should be applied to the \teq{\beta_{\hbox{\fiverm T},e}^2
\ll \omega \ll 1} cases in Equations~(\ref{eq:nrbrems_lim})
and~(\ref{eq:nrinvbrems_lim}) increases monotonically from 1.17 to 1.94
as \teq{\Gamma_e} or \teq{\Gamma_p} vary between 1 to 5, the
representative range of power-law distribution indices.  Since such
factors are of the order of unity, we can assert that Coulomb
correction factors are only marginally important for bremsstrahlung
computations, and will prove immaterial to the considerations and
conclusions of this paper; we neglect them in the algebraic expressions
hereafter.

\subsection{Emission from Ultra-relativistic Species}
 \label{sec:iber}

Hard X-rays and gamma-rays from bremsstrahlung and inverse
bremsstrahlung are generated by ultra-relativistic electrons and ions.
In this regime, a single cross-section cannot be used for the two
mechanisms due to the introduction of aberrations in photon angles and
modifications of energies in transforming between the two rest frames
of the interacting particles.  For normal \teq{e-p} bremsstrahlung, the
differential cross-section is the ultra-relativistic specialization of
the Bethe-Heitler formula (e.g. see Jauch and Rohrlich 1980),
obtained in the Born approximation:
\begin{equation}
   \dover{d\sigma}{d\omega}\biggl\vert_{\hbox{\sixrm BH}}\; =\;
   4\, Z^2\,
   \dover{\fsc\, r_0^2}{\omega} \,
   \biggl\{ 1 + \biggl( \dover{\gamma -\omega}{\gamma} \biggr)^2
              - \dover{2}{3}\, \dover{\gamma -\omega}{\gamma} \biggr\} \,
   \biggl( \log_e \dover{2\gamma (\gamma -\omega )}{\omega}
          - \dover{1}{2} \biggr)
   \quad , \quad \gamma\gg 1\;\; ,
 \label{eq:csect_er}
\end{equation}
and is not subject to significant Coulomb corrections.  Here,
\teq{\gamma} is the electron Lorentz factor.  This formula can be used
to compute contributions to the bremsstrahlung spectrum from mildly
relativistic energies upwards.  The review paper of Blumenthal and
Gould (1970) provides a useful and enlightening discussion of various
features and issues of bremsstrahlung from ultra-relativistic
particles.  From Equation~(\ref{eq:crdists}), assuming a
non-relativistic thermal speed \teq{c\beta_{\hbox{\fiverm T},s}} for a
species \teq{s}, the Lorentz factor distribution is easily deduced to
be
\begin{equation}
   n_s(\gamma ) \; =\;
   \dover{n_s\epsilon_s}{(\beta_{\hbox{\fiverm T},s})^{1-\Gamma_s}}
   \;\dover{3(\Gamma_s-1)}{\Gamma_s-1+3\epsilon_s} \; \gamma^{-\Gamma_s}\;\; .
  \label{eq:crdists_er}
\end{equation}
This can be readily folded with the differential cross-section in
Equation~(\ref{eq:csect_er}) via an adaption of
Equation~(\ref{eq:prodrate}) to yield an emissivity appropriate to the
gamma-ray band:
\begin{equation}
   \dover{dn_{\gamma}(\omega )}{dt}\biggl\vert_{\rm brems}\; \approx\;
   \dover{12 \Lambda\,\epsilon_e}{(\beta_{\hbox{\fiverm T},e})^{1-\Gamma_e}}\,
   \dover{\Gamma_e-1}{\Gamma_e-1+3\epsilon_e}\,
      \kappa (\Gamma_e, \, \omega)\; \omega^{-\Gamma_e}
        \quad  , \quad  1 \,\lesssim\, \omega\; ,
 \label{eq:erbrems_lim}
\end{equation}
where \teq{\Lambda} is given by equation~(\ref{eq:Lambdadef}), and
\begin{eqnarray}
   \kappa (\alpha, \, \omega) & \equiv &
       \int_1^{\infty} \dover{dt}{t^{\alpha}}
   \biggl( \dover{4}{3} - \dover{4}{3t} + \dover{1}{t^2} \biggr)\,
   \Bigl[ \log_e [2\omega t(t-1)] - \dover{1}{2} \Bigr]  \nonumber\\
   & = &
   \dover{3\alpha^2+\alpha + 4}{3(\alpha+1)\alpha (\alpha -1)}\;
   \Bigl[ \log_e 2\omega - \dover{1}{2} - {\cal C} \Bigr]
   + \dover{3\alpha^4 + 2\alpha^3 + 15\alpha^2 - 4}{
            3(\alpha+1)^2\alpha^2 (\alpha -1)^2}
 \label{eq:kappadef} \\
   && - \biggl( \dover{4}{3} \dover{\psi (\alpha -1)}{\alpha -1}
            - \dover{4}{3} \dover{\psi (\alpha )}{\alpha}
            + \dover{\psi (\alpha +1)}{\alpha +1} \biggr)\;\; , \nonumber
\end{eqnarray}
where \teq{\psi (x) = d[\log_e\Gamma (x)]/dx} is the logarithmic
derivative of the Gamma function, and \teq{{\cal C}=\psi
(1)=0.577215\dots} is Euler's constant.  The evaluation of the integral
is facilitated by identity 4.253.6 of Gradshteyn and Ryzhik (1980).
The \teq{\omega\sim 1} values of Equations~(\ref{eq:nrbrems_lim})
and~(\ref{eq:erbrems_lim}) are clearly of comparable magnitude.  Note
also that the bremsstrahlung emissivity traces the electron energy
distribution in the relativistic limit.  A simple derivation of the
form of Eq.~(\ref{eq:erbrems_lim}) is provided in the Appendix.

Determining the emissivity for inverse bremsstrahlung in the limit of
ultra-relativistic ions is, in principal, more involved, largely
because there are no published expressions for the differential
cross-section, integrated over the various interaction angles, that are
general enough to take the place of the Bethe-Heitler formula.  Exact
numerical computations in the Born approximation are presented by Haug
(1972).  Simple analytic formulae for the cross-section were obtained
by Jones (1971), who used the well-known Weizs\"acker-Williams method
(Weizs\"acker 1934; Williams 1935) where the process is treated as
stationary electrons Compton scattering the virtual photons carried by
the proton's electromagnetic field.  Such a technique is quite
applicable to hard X-ray and soft gamma-ray energies, but becomes
erroneous when \teq{\omega\gg 1} since then the approximation of the
Coulomb potential of the proton by a Fourier decomposition into plane
waves (i.e.  mimicking free photons) breaks down.  Notwithstanding, the
approximate formulae of Jones (1971) suffice for the purposes of this
analysis (as will become evident shortly); the \teq{\omega\ll 1} limit
of his Equation~(4) is
\begin{equation}
   \dover{d\sigma}{d\omega}\biggl\vert_{\hbox{\sixrm IB-WW}}\; =\;
   \dover{16}{3}\, Z^2\,
   \dover{\fsc\, r_0^2}{\omega} \,
   \log_e \dover{0.68 \gamma}{\omega}
   \quad , \quad \gamma\gg 1
   \quad , \quad \omega\ll 1\;\; ,
 \label{eq:dsigibww}
\end{equation}
where we have included an extra factor of \teq{Z^2} to include
consideration of ions other than protons.  At photon energies
\teq{1\lesssim \omega\ll\gamma}, the numerical computations in Haug
(1972) indicate that the differential cross-section approximates a
\teq{\omega^{-2}} power-law (exceeding the analytic approximation of
Jones 1971), before rolling over near the kinematic maximum energy of
\teq{\omega\sim\gamma}.  Equation~(\ref{eq:dsigibww}) is readily
integrated over the power-law Lorentz factor distribution for ions that
can be obtained from Equation~(\ref{eq:crdists_er}):
\begin{equation}
   \dover{dn_{\gamma}(\omega )}{dt}\biggl
   \vert_{\hbox{\sevenrm inv}\atop \hbox{\sevenrm brems}}\; \approx\;
   \dover{16 \Lambda\,\epsilon_p}{(\beta_{\hbox{\fiverm T},p})^{1-\Gamma_p}}\,
   \dover{\Gamma_p-1}{\Gamma_p-1+3\epsilon_p}\, \rho (\Gamma_p)\;
        \omega^{-\Gamma_p}
        \quad  , \quad  \omega \,\lesssim\, 1\; ,
 \label{eq:erinvbrems_lim}
\end{equation}
where
\begin{equation}
   \rho (\alpha )\; =\; \int^{\infty}_1 \dover{dt}{t^\alpha} \log_e(0.68t)
     \; =\; \dover{\log_e 0.68}{\alpha -1} + \dover{1}{(\alpha -1)^2}\;\; .
 \label{eq:rhodef}
\end{equation}
This approximation to the contribution to the inverse bremsstrahlung
emissivity from relativistic ions is still correct, up to a factor of
order unity, when \teq{\omega\sim 1}, and hence the specification of
the range \teq{\omega\lesssim 1} in
Equation~(\ref{eq:erinvbrems_lim}).  When \teq{\omega\gtrsim 1}, the
spectrum breaks to assume a \teq{\omega^{-(1+\Gamma_p)}} form,  so that
the inverse bremsstrahlung spectrum is steeper than its bremsstrahlung
counterpart by an index of roughly unity.  We note that while the
emissivity in Equation~(\ref{eq:erinvbrems_lim}) can formally exceed
that in Equation~(\ref{eq:nrinvbrems_lim}), in practice this never
arises due to the distribution index \teq{\Gamma_p} exceeding 2
considerably at near-thermal energies, while \teq{\Gamma_p\lesssim 2}
at relativistic energies.  Hence, Equation~(\ref{eq:nrinvbrems_lim})
should be regarded as the appropriate form for optical and X-ray
energies.  A comparison of Equations~(\ref{eq:erbrems_lim})
and~(\ref{eq:erinvbrems_lim}) again yields the result that the factor
\teq{[\epsilon_p/\epsilon_e]\,(\beta_{\hbox{\fiverm T},p} /
\beta_{\hbox{\fiverm T},e})^{\Gamma_p-1}} roughly defines the ratio of
inverse bremsstrahlung to bremsstrahlung emissivities at soft gamma-ray
energies.

It is also pertinent to briefly mention electron-electron
bremsstrahlung.  Since this process is a quadrupole interaction in a
classical description, it is strongly suppressed for non-relativistic
impact speeds relative to electron-ion bremsstrahlung.  This is borne
out in the differential cross-sections derived by Fedyushin (1952) and
Garibyan (1953), which are of the order of \teq{d\sigma /d\omega \sim
4\fsc r_0^2/(15\omega )} (see, for example, the exposition in the
Appendix of Baring et al.  1999), a factor of \teq{\beta^2} smaller
than the result in Equation~(\ref{eq:csect_nr}).  However, the
\teq{e-e} bremsstrahlung cross-section for ultra-relativistic impact
speeds is necessarily comparable to that for \teq{e-p} bremsstrahlung
(i.e.  Equation~[\ref{eq:csect_er}]), as is indicated in the work of
Baier, Fadin \& Khoze (1967; see also the Appendix of Baring et al.
1999, and the discussion in Blumenthal \& Gould 1970), who derived
limiting forms using the Weizs\"acker-Williams method.  The similarity
of cross-sections for \teq{\gamma\gg 1} implies that the ratio of
electron-electron bremsstrahlung to electron-proton bremsstrahlung is
of the order of unity for plasmas with fairly normal ionic abundances,
a fact that was noted by Baring et al. (1999).  Hence, \teq{e-e}
bremsstrahlung can play an influential role in the determination of the
importance of inverse bremsstrahlung.

\subsection{Expectations for Shock-Heated Environments}
 \label{sec:expect}

Collecting all the rates so far assembled, it is now a simple
enterprise to assert when inverse bremsstrahlung is significant in
astrophysical scenarios.  Setting \teq{{\cal R}_{\rm O-X}} to be the
ratio of Equations~(\ref{eq:nrinvbrems_lim}) to~(\ref{eq:nrbrems_lim})
at \teq{\omega\sim (\beta_{\hbox{\fiverm T},e})^2}, so as to represent
the ratio of inverse bremsstrahlung to bremsstrahlung at optical to
X-ray energies, and \teq{{\cal R}_{\rm MeV}} to be the ratio of
Equations~(\ref{eq:erinvbrems_lim}) to~(\ref{eq:erbrems_lim}) at
\teq{\omega\sim 1}, representing the inverse
bremsstrahlung/bremsstrahlung ratio at soft gamma-ray energies, we
arrive at
\begin{equation}
   {\cal R}_{\rm O-X}\; \sim\; \dover{\epsilon_p}{\epsilon_e}\;
   \biggl( \dover{m_e}{m_p}\, \dover{T_p}{T_e} \biggr)^{(\Gamma_p-1)/2}
   \;\sim\; {\cal R}_{\rm MeV}\;\; ,
 \label{eq:ratios}
\end{equation}
expressing the thermal speeds of the two species in terms of the
electron and proton temperatures \teq{T_e} and \teq{T_p}.  Similar
order-of-magnitude estimates for such ratios can be obtained for other
ionic species such as He$^{2+}$.  In Equation~(\ref{eq:ratios}), it
must be emphasized that for the optical/X-ray band ratio, the index
\teq{\Gamma_p} refers to {\it the mean index ranging from thermal proton
speeds up to thermal electron ones}, while for \teq{{\cal R}_{\rm MeV}},
\teq{\Gamma_p} represents the mean index of protons with \teq{\gamma\gg
1}.  Hence, given the general upward curvature of spectra for
cosmic rays accelerated in non-linear shocks (e.g. Eichler 1984; Jones \& Ellison, 1991), we expect \teq{\Gamma_p >2} for the optical/X-ray band
considerations and \teq{\Gamma_p\lesssim 2} for the gamma-ray band. 
Since the {\it gamma-ray} spectrum for inverse bremsstrahlung (see
the discussion after Equation~[\ref{eq:erinvbrems_lim}]) is steeper
than that for bremsstrahlung in Equation~(\ref{eq:erbrems_lim}),
\teq{{\cal R}_{\rm MeV}} defines an upper bound to the ratio of
emissivities for the two processes in hard gamma-rays, a bound that is
actually conservative given the expectation that \teq{e-e}
bremsstrahlung will amplify the bremsstrahlung signal by a factor
of order 2 in the gamma-ray band.

The ratio \teq{\epsilon_p/\epsilon_e} of efficiency factors is a free
parameter for a shock acceleration model.  As noted at the beginning of
Section~\ref{sec:ibveb}, while efficient acceleration of protons is
generally expected due to strong turbulence local to the shock (e.g.
Ellison, Baring \& Jones 1996), so that \teq{\epsilon_p\sim 0.1}, to
order of magnitude (which is borne out in observations of
interplanetary shocks: see Baring et al. 1997), much less is known
about the value of \teq{\epsilon_e}, essentially due to a paucity of
{\it in situ} observations of electron acceleration in the
heliosphere.  However, this difficulty can be circumvented in part (at
least in astrophysical scenarios) by arguing that cosmic ray data are a
strong indicator of the acceleration efficiency in typical cosmic ray
sources, principally supernova remnant shocks.  The electron-to-proton
abundance ratio above 1 GeV (where both species are relativistic) is
well-known to be of the order of a few percent (e.g. see M\"uller et
al. 1995), which constrains the typical acceleration efficiency,
assuming that differences in electron and proton propagation in the
interstellar magnetic field at these energies are not great.  From the
forms in Eq.~(\ref{eq:crdists}), this ratio can be estimated as a
function of the temperatures, masses and spectral indices of the two
species.  For an isothermal plasma with \teq{\Gamma_e\sim\Gamma_p\sim
2}, one quickly determines that \teq{n_e(\hbox{\rm 1 GeV/c})/n_p(\hbox{
\rm 1 GeV/c}) \sim [\epsilon_p/\epsilon_e]\, (m_e/m_p)^{1/2}}, so that
the observed abundance ratio at momenta 1 GeV/c is reproduced only if
\teq{\epsilon_e} is of the same order of magnitude as \teq{\epsilon_p},
i.e. injection of electrons into the Fermi mechanism is indeed
efficient.  Relaxing the isothermal and equal index stipulations yields
a range of \teq{\epsilon_p/\epsilon_e} on either side of unity.  As
will be apparent in the subsequent subsection, a specific simulation of
the acceleration mechanism applied to supernova remnants yields values
of \teq{\epsilon_p/\epsilon_e} not too disparate from unity.

The temperature ratio \teq{T_e/T_p} clearly controls the importance of
inverse bremsstrahlung relative to bremsstrahlung, both through its
explicit appearance in Eq.~(\ref{eq:ratios}), and also via its
influence on the efficiency ratio \teq{\epsilon_p/\epsilon_e}.  In
shocked plasmas, this ratio is expected to far exceed \teq{m_e/m_p},
thereby implying that {\it inverse bremsstrahlung is insignificant in
most astrophysical scenarios involving shocked plasmas}.  This is the
principal result of this paper; it is contingent upon there being far
fewer protons than electrons above electron thermal speeds, an
occurrence we believe to be common near astrophysical shocks.

Evidence for this can be derived from both observational and
theoretical sources.  On the observational side, there is a clear
indication of the lack of velocity equipartition in electron and proton
populations near the Earth's bow shock.  Feldman (1985) exhibits
measurements of electron and ion velocity distributions that lead to
the identification of (i) strong electron heating during shock passage,
and (ii) downstream velocity thermal ratios for electrons to ions
(mostly protons) of the order of 50 (i.e. \teq{m_eT_p/m_pT_e\sim
0.02}).  The situation for travelling interplanetary shocks is
similar.  While ion distributions indicate shock heating that couples
to the ram pressure differential across such weak shocks (e.g. see
Baring et al. 1997 for Ulysses SWICS data, where \teq{T_p\sim 5\times
10^5}K), measurements of electron populations are generally sparse.
Temperatures of the electron component of the solar wind upstream of
interplanetary shocks is typically in the \teq{3\times
10^4}--\teq{2\times 10^5}K range (Hoang et al. 1995), commensurate with
large scale solar wind averages (Phillips, et al. 1995), and therefore
are a sizeable fraction of the downstream ion velocity dispersion
temperatures in the Ulysses data set.  Hence, in the absence of
any electron heating at the low Mach number interplanetary
shocks encountered by Ulysses, \teq{m_eT_p/m_pT_e\sim 10^{-3} -
10^{-2}} will be obtained.  Such values are evinced in the ISEE data
presented in Feldman (1985), which exhibits only modest heating of
electrons by interplanetary shocks.

Outside the heliosphere, the evidence for significant electron heating
is sparse and more circumstantial.  Shocks at the outer shells of young
supernova remnants (SNRs) provide the best indication.  These move
typically at speeds 500--3000 km/sec, depending on the SNR age and the
mass of the progenitor, suggesting ion heating to temperatures of the
order of \teq{5\times 10^7}--\teq{2\times 10^9}K per nucleon.  Electron
temperatures in SNRs can be deduced from observations of thermal X-ray
emission (for a recent collection of observational studies, see
Zimmermann, Tr\"umper, \& Yorke 1996), and are typically \teq{\sim
10^7}K.  Therefore, it is expected that \teq{m_eT_p/m_pT_e\sim 5\times
10^{-3}}--\teq{0.2} in these sources.  Inferences of electron to
proton temperature ratios in active galactic nuclei are inconclusive,
largely due to the absence of a marker for ion temperatures.

The theoretical expectation that \teq{m_eT_p/m_pT_e\ll 1} is derived
from considerations of dissipation in the shock layer.  In a diffusive
acceleration process where spatial diffusion is effected via {\it
elastic} collisions of particles with magnetic irregularities that are
anchored in the background fluid, the acceleration from thermal
energies is effectively a velocity-dependent process (e.g. see the
reviews of Drury 1983; Jones \& Ellison 1991).  This situation
corresponds to dissipative heating in the shock layer that is
independent of mass of the species.   Baring et al. (1997) illustrates
such a situation in their modelling of accelerated proton and He
populations at nearby interplanetary shocks.  However, dissipation in
the shock layer cannot be entirely elastic in origin in any frame of
reference, due in part to the different mobilities of electrons and
ions.  In a quasi-perpendicular shock, it is clear that the much
smaller Larmor radii of electrons (relative to ions) that cross the
shock lead to the establishment of induced electric fields via charge
separation.  In quasi-parallel shocks, similar electric fields exist,
due to the turbulence of the field structure in the shock layer.  In
fact, generalized shock structure including out-of-the plane magnetic
field components requires the presence of electric fields in all frames
of reference (Jones \& Ellison 1987; 1991).  Such fields heat the low
energy electrons at the expense of the energy of the ions (see the
discussion of Jones 1999), and thereby effect at least a partial
equilibration in energy, providing significant theoretical support for
the contention that \teq{m_eT_p/m_pT_e\ll 1}.  In general, a complex
array of wave modes and instabilities can precipitate strong
acceleration/heating of electrons.  Limited evidence for such heating
can be found in the hybrid plasma simulations of Cargill and
Papadopoulos (1988) for non-relativistic shocks, and the full plasma
simulations of Hoshino et al. (1992) for relativistic shocks in a
pair-dominated ion-pair plasma.  However, these two works do not
simultaneously probe fully three-dimensional hydrogenic plasmas 
with truly realistic ion/electron mass ratios (so as to address a 
full array of wave modes).  Three-dimensional simulations are 
required to accurately account for particle diffusion in all
directions (e.g. see the discussion in Jones, Jokipii \& Baring 1998).
Hence, a definitive simulational assessment of the degree of electron 
heating in shocked plasmas remains an outstanding problem.

\subsection{A Simulational Illustration}
 \label{sec:simul}

The expected dominance of bremsstrahlung can be illustrated using model
predictions from a Monte Carlo simulation of diffusive acceleration in
non-relativistic shocks.  The Monte Carlo technique used to model
diffusive shock acceleration is well-documented in the literature
(e.g.  see Jones \& Ellison 1991; Baring, Ellison, \& Jones  1993;
Ellison, Baring, \& Jones 1996; and more recently, Baring et al. 1999,
for a description of electron injection).  It is a kinematic model,
closely following Bell's (1978) approach to diffusive acceleration,
where the simulation is used to calculate, in effect, solutions to a
Boltzmann equation for particle transport involving a collision
operator, without making any assumption concerning the isotropy of
particle distributions.  Particles are injected at a position far
upstream and allowed to convect into the (infinite plane, steady-state)
shock, diffusing between postulated scattering centers (presumably
magnetic irregularities in the background plasma and self-generated
turbulence) along the way.  As particles diffuse between the upstream
and downstream regions, they continually gain energy in accord with the
Fermi mechanism.  Particle splitting is used very effectively to
maintain excellent statistics over a large dynamic range.  The approach
adopts a phenomenological mean free path, \teq{\lambda_i\propto r_{\rm
g}^{\alpha}} (with \teq{\alpha\sim 1} for all ions and energetic
electrons; for near-thermal electrons, see Baring et al. 1999 for
details) to encompass the complications of plasma microphysics, where
\teq{r_{\rm g} = pc/(ZeB)} is the particle gyroradius.  The aptness of
this power-law prescription to shocked plasma environments is supported
by particle observations at low energies at the Earth's bow shock,
Ellison, M\"obius, \& Paschmann 1990), deductions from ions accelerated
in solar particle events (Mason, Gloeckler, \& Hovestadt 1983), and
also from plasma simulations (Giacalone, Burgess \& Schwartz 1992).
Furthermore, in the strong field turbulence expected (and seen) at
shocks, near-Bohm diffusion, \teq{\lambda_i\gtrsim r_{\rm g}}, imposes
an \teq{\alpha\sim 1} requirement.

%

\centerline{}
\vskip 0.0truein
\centerline{\psfig{figure=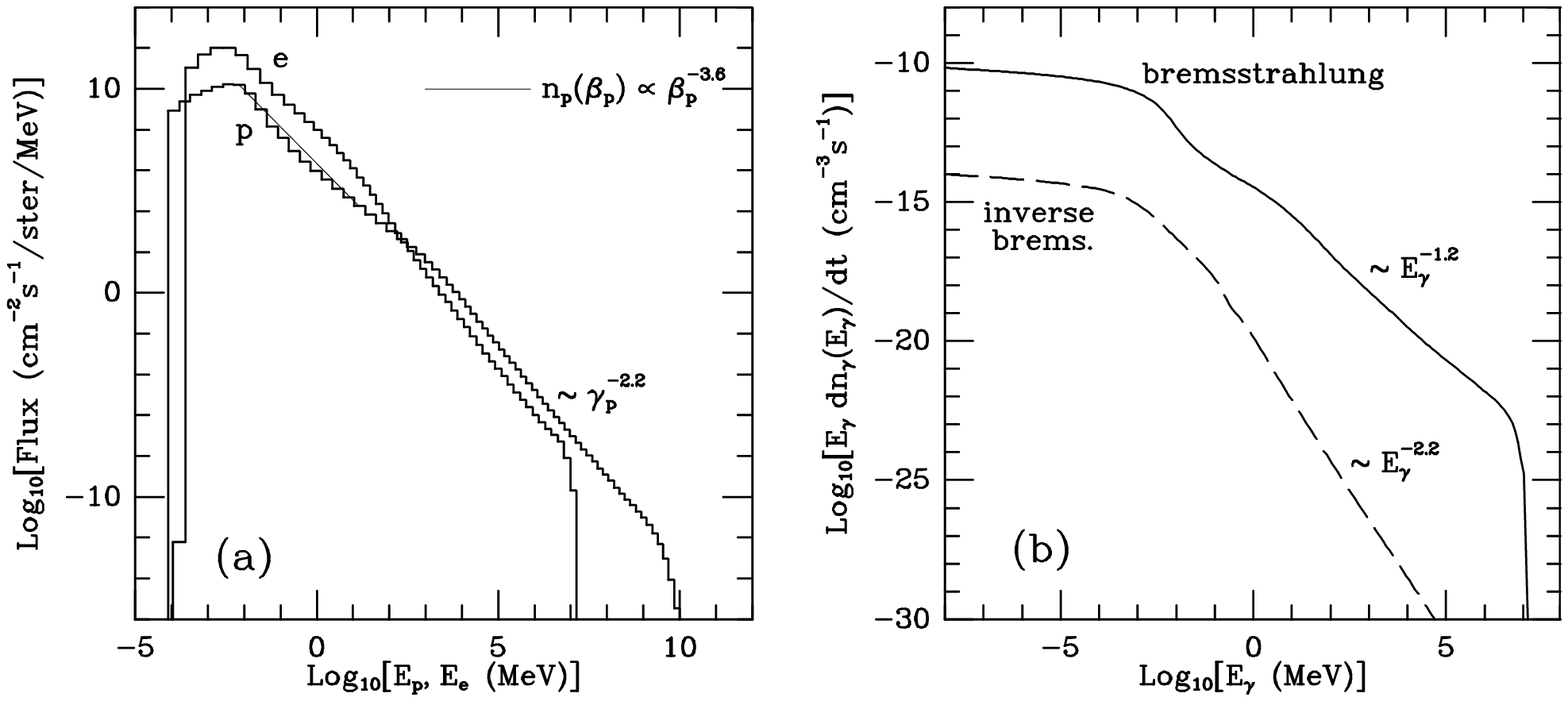,width=18cm}}
\figureout{apj99be_ib_f1.ps}{
Accelerated particle distributions (a) from a Monte Carlo simulation of
acceleration at steady-state cosmic-ray modified non-relativistic
shocks, specifically from the modelling by Ellison, Goret \& Baring
(1999) of broad-band spectra for the supernova remnant Cas A, together
with (b) the bremsstrahlung and inverse bremsstrahlung emissivities
resulting from these distributions.  The electron and proton
distributions are omni-directional fluxes, binned in kinetic energy
intervals, in accord with our previous expositions, and represent the
differential energy distributions multiplied by the particle speed;
they were obtained for a shock speed of 2800 km/sec and an ISM density
of \teq{3}cm$^{-3}$.  An approximate power-law velocity distribution
with index \teq{\Gamma_p=3.6} is depicted to facilitate the discussion
in the text.  The photon emissivities are multiplied by photon energy,
i.e., \teq{E_{\gamma}} times the differential spectrum.
    \label{fig:comparison}
}
\vskip 0.1truein

Typical predictions for the proton and electron distributions
(downstream of a shock) using the Monte Carlo technique are depicted in
Figure~\ref{fig:comparison}, taken from the Ellison, Goret \& Baring's
(1999) study of shock acceleration in the supernova remnant Cas A
(similar spectra are presented in Baring et al. 1999).  The
distributions are exhibited as omni-directional fluxes (i.e.
differential energy fluxes multiplied by particle velocity: \teq{v\,
dn/dE\equiv c\beta n(E)}), mimicking the spectra measured by particle
detectors in the heliosphere.  Hence, for non-relativistic speeds, the
momentum (or velocity) distribution index \teq{\Gamma_s} is exactly
twice the slope of the curves in Figure~\ref{fig:comparison}a, while
for ultra-relativistic particles, \teq{\Gamma_s} is identical to the
depicted slopes, which for both the electrons and protons is roughly
\teq{\Gamma_e\approx\Gamma_p\approx 2.2}.  The particular simulation
output that is illustrated possesses a hot electron component in the
downstream plasma, with a temperature roughly half that of the
protons.  The light power-law drawn in Figure~\ref{fig:comparison}a,
with a velocity distribution as labelled, approximates the proton
distribution between thermal proton and thermal electron speeds; it has
an index of \teq{\Gamma_p\approx 3.6} (i.e. flux distribution
\teq{\beta_p n_p(E_p) \propto E_p^{-1.8}}).  Observe that the Monte
Carlo curves are not pure power-laws in momentum but exhibit a slight
upward curvature in the relativistic portions of the distributions (and
also significant curvature at suprathermal non-relativistic energies
for the protons), a central feature of the non-linear feedback between
the ion diffusion and the shock hydrodynamics in cosmic-ray modified
shocks that is well-documented (e.g. Eichler 1984; Ellison \& Eichler
1984; Jones \& Ellison 1991; Ellison, Jones \& Baring 1996).  Because
such hydrodynamics is strongly regulated by the feedback between
particle acceleration and escape from the remnant, the generated proton
distributions are not very sensitive to the prescription of
\teq{\lambda} as a function of rigidity.

Electrons, on the other hand, are dynamically unimportant, so that
their spectral normalization (but not spectral shape) is sensitive only
to their injection efficiency, which couples to the electron-to-proton
temperature ratio.  While this is somewhat uncertain, we used
reasonable shock speeds and observed X-ray temperatures from remnants
(e.g. Zimmermann, Tr\"umper, \& Yorke 1996) to adopt appropriate
values.  Note also that the ratio of the electron to proton populations
in the 1--10 GeV range (contingent upon the assumed injection
efficiency) for the specific results in Figure~\ref{fig:comparison}a
varies between 15\% and 5\%, somewhat larger than observed \teq{e/p}
cosmic ray ratio in this energy range (e.g. see M\"uller et al. 1995);
\teq{e/p} ratios more consistent with the cosmic ray data are readily
obtained (Baring et al. 1999) by the Monte Carlo simulation.

Figure~\ref{fig:comparison}b depicts the emissivities for
bremsstrahlung (solid curve) and inverse bremsstrahlung (dashed curve)
that result from the particle distributions in
Figure~\ref{fig:comparison}a.  The flat spectra below X-ray energies
correspond to \teq{dn_{\gamma}(\omega )/dt\propto \omega^{-1}},
approximating the low energy results in
Equations~(\ref{eq:nrbrems_lim}) and~(\ref{eq:nrinvbrems_lim}), and
interestingly, close to the index observed in the cosmic X-ray
background below 10 keV (thus motivating the application discussed in
Boldt \& Serlemitsos 1969).  Above the electron thermal energy, the
spectra steepen for both processes, and in the gamma-ray regime, the
inverse bremsstrahlung spectrum (\teq{\propto \omega^{-3.2}\approx
\omega^{-(1+\Gamma_p)}}) is steeper than that of bremsstrahlung
(\teq{\propto \omega^{-2.2}\approx \omega^{-\Gamma_e}}), in accord with
Equation~(\ref{eq:erbrems_lim}) and the discussion following
Equation~(\ref{eq:erinvbrems_lim}).

We note that the steep shoulder in the bremsstrahlung spectrum in the
X-ray band defining the roll-off above the thermal emission could
potentially provide an alternative explanation (Baring et al. 1999) to
synchrotron emission for the non-thermal X-rays seen in the supernova
remnants SN 1006 (Koyama et al. 1995), IC 443 (Keohane et al.~1997) and
Cas A (Allen et al.~1997).  The illustrated case, of index \teq{\sim
2.8} for just over half a decade in energy, lies in between the index
of \teq{\sim 2.3} ASCA obtained for IC 443 and the X-ray observations
of Cas A and SN1006 (\teq{\sim 3}).  While this issue is discussed in
more detail in Ellison, Goret \& Baring (1999), it is important to
remark that inferences of the spectral index of a non-thermal component
in X-ray data are very sensitive to the choice of temperature of a
thermal component and the shape of a non-thermal continuum.
Restrictions to pure power-laws for the continuum, while convenient,
are not really appropriate for either synchrotron radiation turnovers
or bremsstrahlung emission mechanisms, so that conclusions about the
physical mechanism in operation should be guarded.  For example, the
RXTE data in Allen et al.~(1997) on Cas A suggests a spectral break at
\teq{16}keV, steepening from an index \teq{\sim 1.8} to one of
\teq{\sim 3.0}, data that at first sight does not appear inconsistent
with a bremsstrahlung model in the transition region from thermal to
non-thermal energies (though probably with a slightly higher
temperature than the case depicted in Figure~\ref{fig:comparison}).
The critical distinction between synchrotron turnover and
``transition'' bremsstrahlung models therefore lies in the expected
flattening above the transition region in the bremsstrahlung spectrum
(e.g. see Figure~\ref{fig:comparison}b).  This provides an
observational discriminant via broader-band observations beyond the
scope of ASCA and RXTE, though current instrumental sensitivities (such
as those of XMM and Integral) are probably insufficient to answer this
question.  A related situation arises with diffuse/unresolved emission
in the galactic ridge, and recent broad-band X-ray spectra of the
Scutum arm (Valinia, Kinzer \& Marshall 1999) suggest a turnover in the
50--100 keV range, which could indicate the superposition of
Comptonized (or synchrotron) emission from unresolved discrete
sources.  However, we note that data statistics in the OSSE band are
not sufficient to preclude a flattening at higher energies that would
be characteristic of bremsstrahlung of either discrete source or
diffuse origin.

Returning to the issue at hand, the particular example in
Figure~\ref{fig:comparison} gives a strong dominance of bremsstrahlung
over inverse bremsstrahlung.  In fact, given the \teq{\Gamma_p\approx
3.6} index for non-relativistic speeds, and \teq{T_e/T_p\sim 1/2},
Equation~(\ref{eq:ratios}) can be simply applied to estimate a ratio of
\teq{\sim 8000} between the two processes (for
\teq{\epsilon_p/\epsilon_e=1}), close to the ratio deduced in the
optical band from Figure~\ref{fig:comparison}b.  This agreement
highlights the viability and usefulness of the schematic distributions
in Eq.~(\ref{eq:crdists}), and also that of the consequent quantitative
estimates of the various emissivities throughout this paper.  Observe
that the only way to increase \teq{{\cal R}_{\rm O-X}} is to
dramatically lower the temperature of the electrons relative to the
protons, but to do so would cause a corresponding decrease in the
\teq{e/p} ratio at relativistic energies (and influence the ratio
\teq{\epsilon_p/\epsilon_e} of injection efficiencies), well below
observed values in the 10 GeV band (see, e.g. M\"uller et al. 1995, for
a recent compendium of data).  As argued above, it is unlikely that
\teq{{\cal R}_{\rm O-X}\gtrsim 0.01} will be realized often in shocked
astrophysical plasmas.

It is natural to question whether such a result holds for ions heavier
than protons.  Clearly, both the bremsstrahlung and inverse
bremsstrahlung contributions from helium and heavier elements are given
by their abundance (usually small) multiplied by \teq{Z^2}.  The
\teq{Z^2} enhancement provides significant bremsstrahlung emissivities
when helium targets (typically of \teq{\sim 5\%} abundance) are
involved.  For inverse bremsstrahlung, the \teq{Z^2} amplification
potentially could be offset by a more severe mass ratio reduction, if
lower thermal speeds are precipitated downstream of shocks.  There is
little or no evidence for such lower thermal speeds for ions heavier
than protons.  In fact, the Ulysses SWICS data on the 91097
interplanetary shock unequivocally indicates the contrary: the shock
dissipation mechanism generates virtually identical velocity
dispersions for both protons and alpha particles (Baring, et al.
1997).  Such a property is well-modelled by the Monte Carlo technique,
since the elastic scattering hypothesis that it employs renders the
shock acceleration model a velocity-dependent injector.  This
scattering assumption, together with a rigidity-dependent diffusion
coefficient, is rigorously put to the test in the successful prediction
(Meyer, Drury, \& Ellison 1997; Ellison, Drury, \& Meyer 1997) of
abundances of ions of various metallicities in the cosmic ray
population, using a supernova remnant site for the acceleration of dust
grains.  These pieces of evidence point to an approximate velocity
equipartition of thermal ions in shocks, so that the criterion in
Equation~(\ref{eq:ratios}) renders the conclusion that inverse
bremsstrahlung is generally insignificant in shocked astrophysical
plasmas also applicable to contributions from ions heavier than protons.

\section{DISCUSSION}
 \label{sec:discuss}

The unimportance of inverse bremsstrahlung for shocked environs in
discrete sources, the principal result of this paper, can be
interpreted in the broader context of invocations of inverse
bremsstrahlung in diffuse emission problems.  The pre-eminent implication
of this result is obviously that supernova remnants will dominate their
environs in X-ray and gamma-ray luminosity, and therefore will obscure
truly diffuse emission components in extended regions such as the
galactic ridge and the Orion complex, if the source remnants are too
densely clustered.  Hence, it is of significant interest to identify the
length scales on which discrete sources such as supernova remnants
provide a dominant contribution to unresolved X-ray and gamma-ray
emission in the galaxy, as a guide to modelers and also to aid the
interpretation of experimental results.

The discussion of this issue is not restricted to just the ratio of
bremsstrahlung from cosmic ray primary (i.e. shock-accelerated)
electrons and inverse bremsstrahlung from cosmic ray ions.  The works
of Tatischeff, Ramaty \& Kozlovsky (1998) and Dogiel et al. (1998)
argued that suprathermal proton bremsstrahlung emission levels are
comparable to those of bremsstrahlung from knock-on electrons in the
Orion complex, given the then (and now defunct) Comptel nuclear line
detection in this region.  Knock-on electrons are generated from the
cool ambient interstellar population via Coulomb collisions with
energetic cosmic ray ions (and perhaps electrons).  Since the stopping
length (on ambient electrons {\it or} protons) for 25 keV electrons in
interstellar environments of typical density is of the order of 100 pc
(e.g. see Valinia \& Marshall 1998), Coulomb collisions are clearly
relevant for electrons in large scale systems such as the Orion cloud
complex.  Coulomb scattering simultaneously mediates both energization
of ambient electrons and cooling of cosmic ray primary electrons.  In
contrast, shocked astrophysical plasmas such as those in young to
middle-aged supernova remnants are collisionless, so that knock-on
populations are locally negligible.  Given the conclusions of this
paper relating to shock-heated plasmas, it is salient to also examine
when knock-on electron bremsstrahlung is a significant contributor to
diffuse/unresolved emission relative to primary electron
bremsstrahlung.  Note that the comparisons of Tatischeff, Ramaty \&
Kozlovsky (1998) and Dogiel et al. (1998) of inverse bremsstrahlung
relative to knock-on electron bremsstrahlung did not include an
abundant, shock-accelerated electron population, and therefore did not
address this issue.

\clearpage

\subsection{Knock-on Electrons}
 \label{sec:knockon}

First, we address the question of
whether knock-on electrons are abundant relative to
cosmic ray primary electrons.
For mono-energetic cosmic rays, non-relativistic knock-on electrons
possess an \teq{E^{-2}} kinetic energy distribution regardless of
whether the energetic cosmic rays are ions (e.g. see Hayakawa 1969) or
electrons (Gould 1972; true also for relativistic electrons:  see
Baring 1991), which corresponds to a \teq{\beta^{-3}} velocity
distribution.  This is comparable in steepness to the shock-accelerated
electron distribution exhibited in Fig.~\ref{fig:comparison}a.
However, in spite of the ambient (and therefore also knock-on)
electrons being much more abundant than cosmic ray ones, the relative
population level at almost relativistic speeds defines the criterion
for the importance of knock-on electrons in producing X-ray
bremsstrahlung.  The knock-on electron population extends upward from
extremely low thermal energies (temperatures of 3--100 K; see Dogiel et
al. 1998 for a listing of parameters for the Orion complex).  Assuming
that only a minority of ambient electrons can be energized in knock-on
collisions without disturbing the thermal balance of the ambient
interstellar medium, it is straightforward to determine that the
distribution \teq{n_e(E_{\hbox{\sevenrm MeV}})\sim 10^{-9} n_e\,
T_{1}\, E_{\hbox{\sevenrm MeV}}^{-2}} MeV$^{-1}$ cm$^{-3}$ forms a
conservative upper bound to the knock-on energy distribution out to MeV
energies.  Here \teq{T_{1}} is the ISM temperature in units of 10 K,
and \teq{n_e} is the ISM density in units of cm$^{-3}$, typically of
the order of \teq{1} cm$^{-3}$ in the normal ISM, and \teq{\sim 100}
cm$^{-3}$ in dense interstellar regions.  Also, \teq{E_{\hbox{\sevenrm MeV}}}
is the electron energy in units of MeV.  

\centerline{}
\vskip 0.3truein
\centerline{\psfig{figure=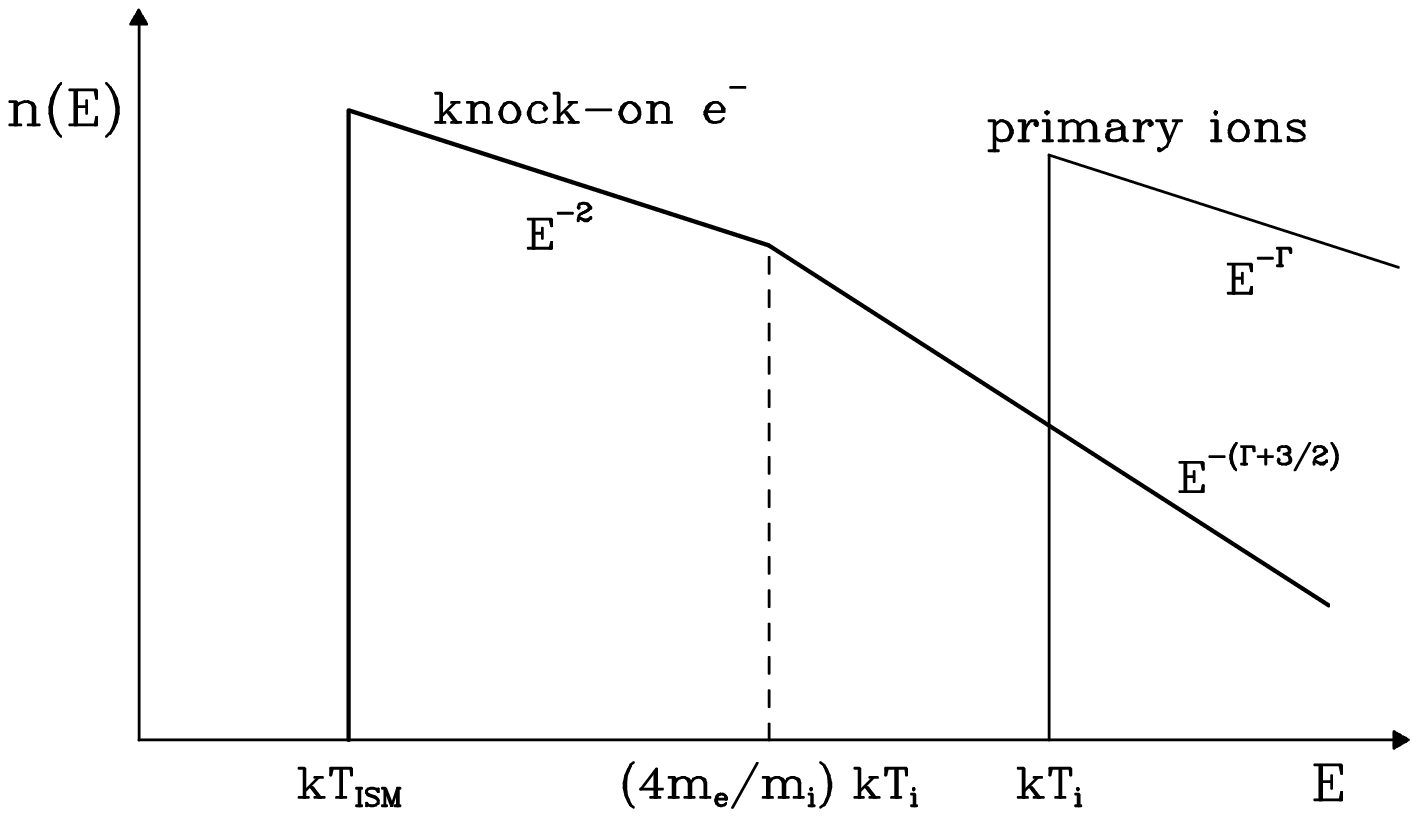,width=14cm}}
\figureout{apj99be_ib_f2.ps}{
A schematic depiction of the knock-on electron distribution (heavy
line) generated by a cosmic ray ion distribution that is a power-law
above thermal energies represented by a non-relativistic temperature
\teq{T_i} (\teq{\ll m_ic^2/k}).  For ions originating in SNRs,
typically \teq{kT_i\sim 10^{-2}}--\teq{1}MeV from shock heating.  The
knock-on distribution extends upwards from the temperature
\teq{T_{\hbox{\fiverm ISM}}} of the ambient ISM, and steepens above the
kinematic break at twice the cosmic ray ion thermal speed (i.e.
\teq{E\sim (4m_e/m_i)kT_i}; see Baring 1991, for example, for a
discussion of kinematics).  While knock-on electrons may dominate
cosmic ray primaries by number, they suffer a substantial paucity at
high energies.
    \label{fig:knock-on}
}
\vskip 0.1truein

This upper bound is actually a gross overestimate, given that it
corresponds to 100\% of the ambient electron population being
energized.  Furthermore, due to the kinematic maxima for the energies
imparted to cold electrons by Coulomb collisions, above the thermal
speed of the ballistic cosmic rays, which for ionic projectiles is
unlikely to exceed typical shock speeds for supernova remnants of a
500-3000 km/sec, the knock-on spectrum steepens from the dependence
\teq{E^{-2}}.  The knock-on distribution can then be shown to roughly
follow an \teq{E^{-3/2-\Gamma_p}} distribution for a cosmic ray ion
spectrum \teq{\propto E^{-\Gamma_p}} (derivable, for example, using
Eq.~(9) of Baring 1991), rendering the knock-on electron population at
around 100 keV far smaller than the above bound; see
Figure~\ref{fig:knock-on} for a schematic depiction.  An approximate
refined estimate for the knock-on distribution would then be
\teq{n_e(E_{\hbox{\sevenrm MeV}})\sim 0.1 n_e\, T_{1}\,
[E_{\hbox{\sevenrm MeV}}/10^{-4}]^{-3/2-\Gamma_p}} MeV$^{-1}$ cm$^{-3}$
for \teq{E_{\hbox{\sevenrm MeV}}\gtrsim 10^{-4}}.  This can be compared
with a rough estimate for the non-thermal electron distribution
expected for efficient acceleration at a remnant's shock:
\teq{n_e(E_{\hbox{\sevenrm MeV}})\sim 10^5 n_e\, T_{7}\,
[E_{\hbox{\sevenrm MeV}}/10^{-4}]^{-2}} MeV$^{-1}$ cm$^{-3}$, where here
\teq{T_{7}} is the shocked plasma temperature in units of \teq{10^7}K.  
Since \teq{T_{7}\sim 1} near SNR shocks, clearly the primary cosmic ray
source spectrum locally swamps the knock-on one.  Away from such
shocks, diffusion dilutes the primary source spectrum down to cosmic
ray flux levels.

While a specific prediction for the ambient cosmic ray primary electron
density could be made via a particular acceleration model (such as the
Monte Carlo technique used above), combined with a specific diffusive
propagation model, we prefer to make use of existing experimental
cosmic ray data under the premise that dense interstellar regions are
not extraordinarily peculiar in their cosmic ray properties.  M\"uller
et al. (1995; see also M\"uller \& Tang 1987) exhibit the primary
electron flux at Earth above around 3 GeV.  From their data, one
quickly arrives at the differential energy distribution for electrons
of \teq{n_e(E_{\hbox{\sevenrm MeV}})\sim 10^{-4} \, E_{\hbox{\sevenrm
MeV}}^{-3.25}} MeV$^{-1}$ cm$^{-3}$ at energies above 3 GeV.  Direct
extrapolation of this down to 100 keV (i.e. well-below energies where
heliospheric modulation impacts measurements of the cosmic ray
spectrum) would lead to the conclusion that cosmic ray electrons should
dominate the maximum possible density of knock-on electrons by around 4
orders of magnitude, and the refined estimate above of the knock-on
density by a factor of \teq{10^9} or more (for \teq{\Gamma_p\gtrsim
2}), for realistic values of the ISM temperature \teq{T_1}.  However,
the primary electron spectrum is probably much flatter in the 100 keV
to 3 GeV band than measured at higher energies.  The diffuse radio
synchrotron emission in the 30--600 MHz band (\teq{\propto \nu^{-2/3}})
in our galaxy suggests (e.g. see the review by Webber 1997) a cosmic
ray electron distribution \teq{\propto E^{-7/3}} between 500 MeV and
2.2 GeV.  This flattening probably reflects the transition from
directly sampling contributions to the cosmic ray spectrum near
discrete sources (which generally should possess flatter \teq{\sim
E^{-2}} distributions like those in Figure~\ref{fig:comparison}) to
larger-scale steepening imposed by cosmic ray propagation effects.
Even using an extrapolation \teq{n_e(E_{\hbox{\sevenrm MeV}})\sim
10^{-6} \, E_{\hbox{\sevenrm MeV}}^{-2.67}} MeV$^{-1}$ cm$^{-3}$ at
energies below 300 MeV based on the inferences from the radio data, one
arrives at the conclusion that the cosmic ray primary electron
population would far exceed any knock-on component generated by cosmic
ray ions.

\subsection{Diffuse and Unresolved Emission}
 \label{sec:diffuse}

Hence, it follows that a knock-on population of electrons can only
generate a significant amount of electron bremsstrahlung emission in
dense interstellar regions provided that there is a profound paucity of
cosmic ray electrons in the 30 keV--1 MeV range.  Cooling of electrons
through Coulomb and ionization losses potentially can create such a
paucity; lower energy electrons cool faster than more energetic ones
do.  Therefore, a depletion of cosmic ray electrons below 1 MeV is a
distinct possibility on scales considerably greater than 10pc outside
the supernova remnants that act as sources of cosmic rays.
Specifically, using the numbers quoted by Valinia and Marshall (1998),
25 keV electrons that would emit bremsstrahlung in the RXTE band have a
collisional loss pathlength of around 100pc in the ISM.  This linear
scale \teq{\lambda_{\rm Coul}} can be translated into a radial scale
\teq{r_{\rm Coul}} by incorporating the effects of diffusion due to the
turbulent interstellar field.  The mean free path
\teq{\lambda_{\hbox{\fiverm B}}} for such diffusion is extremely
uncertain, but as is the case for the undisturbed interplanetary medium
in the heliosphere, estimates for \teq{\lambda_{\hbox{\fiverm B}}} (by
coupling to the field irregularity scale; e.g. see Kim, Kronberg \&
Landecker 1988) far exceed particle gyroradii, at around 100pc.  This
implies that for 25 keV electrons, \teq{r_{\rm Coul} = [\lambda_{\rm
Coul}\,\lambda_{\hbox{\fiverm B}}]^{1/2}\sim 100}pc defines the maximal
extent of their ``Coulomb halo'' around a SNR.  For higher energy
electrons, the radius of this halo is larger, so that at any given
radius, the primary electron spectrum is depleted up to the energy at
which Coulomb losses become insignificant.

Irrespective of whether and on what scales such depletions
exist (no data pertaining to cosmic ray electrons samples this energy
range), the dominance of bremsstrahlung from primary cosmic ray
electrons over inverse bremsstrahlung from primary cosmic ray protons
that is established in this paper would lead to the major contribution
to unresolved ``diffuse'' X-ray emission in Orion or the galactic ridge
coming from discrete sources, unless the volume filling factor of
supernova remnants in the interstellar medium is less than
\teq{10^{-4}}--\teq{10^{-3}}.  While accurate determination of this
factor is difficult, extant compendia of supernova remnants (e.g.
Helfand et al.  1989; Green, 1999) lead to estimates at the lower end
of this range (i.e. \teq{\sim 10^{-4}}), to order of magnitude.  Hence,
while it is possible that such conditions might prevail in some
galactic interstellar environments, and more probably at moderate
latitudes above the disk, we contend that primary cosmic ray electron
bremsstrahlung will simultaneously dominate bremsstrahlung from
secondary knock-on electrons and inverse bremsstrahlung mediated by
cosmic ray ions as contributors to diffuse X-ray emission in the most
important and interesting environments.

When this is not the case, primary electron bremsstrahlung signals (of
spectral index 2--3) will be confined to discrete source locales of
scales \teq{\sim 10}--100pc, i.e. near supernova remnants, while
suprathermal proton bremsstrahlung and knock-on electron bremsstrahlung
(of similar spectral indices) will sample the length scales appropriate
to ionic collisional losses, namely \teq{\sim 0.2}Mpc for the 50MeV
protons that can generate inverse bremsstrahlung X-rays in the RXTE
band.  To reiterate, these length scales will be appropriately
contracted in linear dimensions if they exceed the diffusive scales for
such particles in the interstellar magnetic field (e.g. to \teq{\sim
5}kpc from the protons: see Valinia and Marshall 1998).  Such radiation
``clumping'' issues should form a focus of discussions of diffuse and
unresolved emissions.  Note that because \teq{\sim 50}MeV cosmic ray
protons are not easily depleted on scales \teq{\ll 5}kpc, and since the 
low ISM temperature forces the knock-on distribution to low levels, we
expect that generally inverse bremsstrahlung dominates knock-on
electron bremsstrahlung outside remnants.


An observationally interesting question concerns how easily
bremsstrahlung signals from primary cosmic ray electrons associated
with SNRs can by distinguished from truly compact sources such as
neutron stars or black holes, given current angular resolution
capabilities of X-ray telescopes.  Spectral signatures may provide the
answer prior to improvements in instrumental resolving power.  The
similarity of unresolved emission spectra in source-rich and
source-poor galactic regions (ASCA data: Tanaka, Miyaji \& Hasinger
1999), and the spectral similitude of observations of the Scutum arm
(RXTE/OSSE data:  Valinia, Kinzer \& Marshall 1999) and galactic X-ray
binaries and black hole candidates provide clues impinging upon this
issue.

Observe also that the expected general dominance of bremsstrahlung from
primary electrons returns one to the energetics issues raised by Skibo,
Ramaty \& Purcell (1996), who discerned that the general inefficiency of
bremsstrahlung processes in the X-ray band relative to collisional and
ionization losses elicits difficulties in generating the requisite
luminosity.  In passing, we note that clumping of emission in the
proximity of remnants will help alleviate the energy supply problem for
a bremsstrahlung origin of the galactic diffuse X-ray emission, by
diminishing such losses for the electrons.

\section{CONCLUSION}
 \label{sec:conclusion}

In conclusion, the results of this paper indicate that inverse
bremsstrahlung in the optical to X-ray and gamma-ray bands can be
safely neglected in most models of discrete sources invoking shock
acceleration.  This follows from the approximate ratio, posited in
Equation~(\ref{eq:ratios}), of inverse bremsstrahlung to bremsstrahlung
from primary non-thermal electrons, together with expectations for
dissipational heating of electrons in the shock layer.  Both
observational and theoretical evidence favors electron temperatures
almost comparable to, and certainly not very deficient relative to,
proton temperatures in shocked plasmas.  Hence shocked environments in
discrete sources such as supernova remnants should have inverse
bremsstrahlung contributing only in a minor capacity to the overall
emission, a situation that could apply also to larger, extended
interstellar regions such as the galactic ridge, provided that the
volume filling factor associated with remnants exceeds around
\teq{10^{-4}}--\teq{10^{-3}}.  Since this filling factor may sometimes
be lower, perhaps at moderate to high galactic latitudes, we find that
in such cases, inverse bremsstrahlung can contribute significantly in
the X-ray band on length scales of the order of 50--100pc or greater,
thereby posing an interesting observational goal to resolve bright
bremsstrahlung X-ray emission concentrated around discrete sources from
any truly diffuse inverse bremsstrahlung component that spatially
traces the cosmic ray ion population.  We also contend that
bremsstrahlung from knock-on electrons is probably always a minor
contribution, and find that the dominance of bremsstrahlung from
primary cosmic ray electrons over inverse bremsstrahlung in the
vicinity of remnants is even more enhanced in the gamma-ray band.


\acknowledgments
We thank Stephen Reynolds, Elihu Boldt, Azita Valinia and Peter
Serlemitsos for discussions, Vincent Tatischeff and the anonymous
referee for comments helpful to the improvement of the presentation,
and Bob Gould for pointing out certain literature on bremsstrahlung to
us.

\appendix
\section{} 
 \label{sec:appendix}
\vskip -10pt

In this brief Appendix, a pedagogical alternative to the derivation of
the various bremsstrahlung emissivities is provided, using approximate,
simplified cross-sections, so as to elucidate the results obtained in
the text by diminishing the mathematical complexity.  To this end, the
expressions obtained here are proportionalities, possessing only the
principal variables.  Eqs.~(\ref{eq:prodrate}) and~(\ref{eq:crdists})
are again the starting points.  Instead of using
Eqs.~(\ref{eq:csect_nr}) and~(\ref{eq:csect_er}) for the
non-relativistic and ultrarelativistic bremsstrahling differential
cross-sections, it is expedient to use a single approximation
\begin{equation}
   \dover{d\sigma}{d\omega}\biggl\vert_{\hbox{\sixrm NR-ER}}\; \sim\;
   \dover{16}{3}\, Z^2\,
   \dover{\fsc\, r_0^2}{\omega} \,\dover{1}{\beta^2}\,
   \Theta \Bigl( \dover{\omega}{\gamma -1} \Bigr) \quad , 
 \label{eq:csect_nr_approx}
\end{equation}
where \teq{\Theta (x)} is a step function that is unity for
\teq{0<x<1}, and zero otherwise.  Consider first normal
bremsstrahlung.  The integrations over the electron distribution
are straightforward, with both the \teq{p\leq p_{\hbox{\fiverm T},e}}
and \teq{p > p_{\hbox{\fiverm T},e}} portions contributing for
\teq{\omega \lesssim \beta_{\hbox{\fiverm T},e}^2 \ll 1}, and
just the high momentum tail being sampled for 
\teq{\beta_{\hbox{\fiverm T},e}^2 \ll \omega \ll 1}.  The resulting
proportionalities can then be summarized as (for
\teq{\Lambda = Z^2 n_pn_e \fsc r_0^2c})
\begin{equation}
   \dover{dn_{\gamma}(\omega )}{dt}\biggl\vert_{\rm brems}\; \propto\;
   \dover{\Lambda}{\beta_{\hbox{\fiverm T},e}} \,
   \cases{ \dover{1}{\omega} \,
          \Bigl( \dover{1}{2} + \dover{\epsilon_e}{\Gamma_e}\Bigr)\, , &
               $\omega \lesssim \beta_{\hbox{\fiverm T},e}^2 \ll 1$, \cr
          \epsilon_e\; 
          \dover{\beta_{\hbox{\fiverm T},e}^{\Gamma_e}}{
             \omega^{1+\Gamma_e/2}}\,  \vphantom{\Biggl(}\; , &
               $\beta_{\hbox{\fiverm T},e}^2 \ll \omega \ll 1$,        }
 \label{eq:nrbrems_lim_prop}
\end{equation}
dependences that are clearly evident in Eq.~(\ref{eq:nrbrems_lim}); the
absence of logarithmic factors is an obvious consequence of the
simplification of the cross-section.  There is little need to
explicitly state the equivalent proportionality for the case of
non-relativistic inverse bremsstrahlung, which reproduces the principal
components of Eq.~(\ref{eq:nrinvbrems_lim}) in a similar manner.  For
normal bremsstrahlung from ultrarelativistic electrons, the integration
can be performed with equal efficacy using Eq.~(\ref{eq:crdists_er}),
yielding
\begin{equation}
   \dover{dn_{\gamma}(\omega )}{dt}\biggl\vert_{\rm brems}\; \propto\;
   \dover{\Lambda\,\epsilon_e}{(\beta_{\hbox{\fiverm T},e})^{1-\Gamma_e}}\,
   \omega^{-\Gamma_e}
        \quad  , \quad  1 \,\lesssim\, \omega\; ,
 \label{eq:erbrems_lim_approx}
\end{equation}
thereby reproducing the form of Eq.~(\ref{eq:erbrems_lim}), minus the
complexity of integrations over the combinations of logarithmic and
power-law factors.

%
%




\begin{references}
%
\reference{}
   Allen, G.~E. et al. 1997, \apjl\vol{487}{L97}
\reference{}
   Baier, V.~N., Fadin, V.~S. \& Khoze, V.~A. 1967, \jetp\vol{24}{760}
\reference{}
   Baring, M.~G., Ellison, D.~C. \& Jones, F.~C. 1993, \apj\vol{409}{327}
\reference{}
   Baring, M.~G., Ogilvie, K.~W., Ellison, D.~C., \& Forsyth, R.~J. 1997,
   \apj\vol{476}{889}
\reference{}
   Baring, M.~G., Ellison, D.~C., Reynolds, S.~P., Grenier, I.~A., \&
   Goret, P. 1999, \apj\vol{513}{311}
\reference{}
   Bell, A. R. 1978, \mnras\vol{182}{147}
\reference{}
   Boldt, E. 1987, Phys. Rep. \vol{146}{215}
\reference{}
   Boldt, E. \& Serlemitsos, P. 1969, \apj\vol{157}{557}
\reference{}
   Bethe, H.~A. \& Heitler, W. 1934, Proc. Roy. Soc. \vol{A146}{83}
\reference{}
   Bloemen, H., et al. 1994, \aap\vol{281}{L5}
\reference{}
   Bloemen, H., et al. 1999, to appear in Proc. 3rd Integral Workshop.
\reference{}
   Blumenthal, G.~R. \& Gould, R.~J. 1970, \rmp\vol{42}{237}
\reference{}
   Brown, R.~L. 1970, \apjl\vol{159}{L187}
\reference{}
   Cargill, P.~J. \& Papadopoulos, K. 1988, \apjl\vol{329}{L29}
\reference{}
   Dogiel, V.~A., et al. 1998, \pasj\vol{50}{567}
\reference{}
   Drury, L.~O'C. 1983, Rep. Prog. Phys. \vol{46}{973}
\reference{}
   Eichler, D. 1984, \apj\vol{277}{429}
\reference{}
   Ellison, D.~C., Baring, M.~G. \& Jones, F.~C. 1996, \apj\vol{473}{1029}
\reference{}
   Ellison, D.~C., Drury, L. O'C., \& Meyer, J.-P. 1997, \apj\vol{487}{197}
\reference{}
   Ellison, D.~C., \& Eichler, D. 1984, \apj\vol{286}{691}
\reference{}
   Ellison, D.~C., Goret, P., \& Baring, M.~G. 1999, in preparation.
\reference{}
   Ellison, D.~C., M\"obius, E., \& Paschmann, G. 1990, \apj\vol{352}{376}
\reference{}
   Elwert, G. 1939, Ann. Physik \vol{34}{178}
\reference{}
   Fabian, A.~C. \& Barcons, X. 1992, Ann. Rev. Astr. Astrophys.
   \vol{30}{429}
\reference{}
   Fedyushin, B.~K. 1952, \rm Zhur. Eksp. Teor. Fiz. \vol{22}{140}
\reference{}
   Feldman, W.~C. 1985, in Collisionless Shocks in the Heliosphere:
   Reviews of Current Research, Geophys. Monogr. Ser., 35, eds. B.~T.
   Tsurutani and R.~G. Stone (AGU, Washington, DC), p.~195.
\reference{}
   Garibyan, G.~M. 1952, Zhur. Eksp. Teor. Fiz. \vol{24}{617}
\reference{}
   Giacalone, J., Burgess, D. and Schwartz, S. J. 1992, in Proc. 26th ESLAB
   Symposium, \rm Study of the Solar-Terrestrial System \rm (ESA, Noordwijk)
   p.~65
\reference{}
   Gould, R.~J. 1970, Am. J. Phys. \vol{38}{189}
\reference{}
   Gradshteyn, I.~S. and Ryzhik, I.~M. 1980, Table of Integrals, Series
     and Products, (Academic Press, New York)
\reference{}
   Green, D.~A. 1998, A Catalogue of Galactic Supernova Remnants,
   (September 1998 edition) 
   {\it http://www.mrao.cam.ac.uk/surveys/snrs/}
\reference{}
   Haug, E. 1972, Astrophys. Lett. \vol{11}{225}
\reference{}
   Haug, E. 1975, Z. Naturforsch. \vol{30a}{1099}
\reference{}
   Hayakawa, S. 1969, Prog. Theor. Phys. \vol{41}{1594}
\reference{}
   Hayakawa, S. \& Matsuoka, M. 1964, Prog. Theor. Phys. Suppl. \vol{30}{204}
\reference{}
   Helfand, D.~J., Velusamy, T., Becker, R.~H. \& Lockman, F.~J. 1989,
   \apj\vol{341}{151}
\reference{}
   Hoang, S., Lacombe, C., Mangeney, A., Pantellini, F., Balogh, A.,
   Bame, S.~J., Forsyth, R.~J. and Phillips, J.~L. 1995,
   \rm Adv. Space Sci. \rm \vol{15\, (8/9)}{371}
\reference{}
   Hoshino, M., Arons, J., Gallant, Y.~A., \& Langdon, A.~B.
   1992, \apj\vol{390}{454}
\reference{}
   Jauch, M.~M. \& Rohrlich, F. 1980, \rm The Theory of Photons and
   Electrons, \rm (2nd edn. Springer, Berlin)
\reference{}
   Jones, F.~C. 1971, \apj\vol{169}{503}
\reference{}
   Jones, F.~C. 1999, to appear in proc. of the Erice workshop on
   Cosmic Rays, eds. M. Shapiro et al., held in Erice, July 1998.
\reference{}
   Jones, F.~C. \& Ellison, D.~C. 1987, \jgr\vol{92}{11,205}
\reference{}
   Jones, F.~C. \& Ellison, D.~C. 1991, \ssr\vol{58}{259}
\reference{}
   Jones, F.~C., Jokipii, J.~R. \& Baring, M.~G. 1998, \apj\vol{509}{238}
\reference{}
   Keohane, J.~W., Petre, R., Gotthelf, E.~V., Ozaki, M., \& Koyama, K.
   1997, \apj\vol{484}{350}
\reference{}
   Kim, K.-T., Kronberg, P.~P. \& Landecker, T.~L. 1988, \aj\vol{96}{704}
\reference{}
   Koyama, K. et al. 1995, \nat\vol{378}{255}
\reference{}
   Marshall, F.~E., et al. 1980, \apj\vol{235}{4}
\reference{}
   Mason, G.~M., Gloeckler, G., and Hovestadt, D. 1983, \apj\vol{267}{844}
\reference{}
   Maxon, M.~S. \& Corman, E.~G. 1967, \pr\vol{163}{156}
\reference{}
   Meyer, J.-P., Drury, L. O'C., \& Ellison, D.~C. 1997,
   \apj\vol{487}{182}
\reference{}
   M\"uller, D., \& Tang, K.-K. 1987, \apj\vol{312}{183}
\reference{}
   M\"uller, D., et al. 1995, \rm Proc. 24th ICRC (Rome)\rm , \vol{3}{13}
\reference{}
   Pohl, M. 1998, \aap\vol{339}{587} 
\reference{}
   Phillips, J.~L., Bame, S.~J., Gary, S.~P., Gosling, J.~T.,
   Scime, E.~E. \& Forsyth, R.~J. 1995, \ssr\vol{72}{109}
\reference{}
   Skibo, J.~G., Ramaty, R. \& Purcell, W.~R. 1996, \aaps\vol{120}{403}
\reference{}
   Sommerfeld, A. 1931, Ann. Physik \vol{11}{257}
\reference{}
   Tanaka, Y., Miyaji, T.\& Hasinger, G. 1999, Astron. Nachr., in press.
\reference{}
   Tatischeff, V., Ramaty, R. \& Kozlovsky, B. 1998, \apj\vol{504}{874}
\reference{}
   Tatischeff, V., Ramaty, R. \& Valinia, A. 1999, to appear in 
   "LiBeB, Cosmic Rays and Gamma-Ray Line Astronomy", 
   ASP Conference Series, eds. Ramaty, R., et al. [astro-ph/9903326]
\reference{}
   Valinia, A., Kinzer, R.~L. \& Marshall, F.~E. 1999, \apjl\ submitted.
\reference{}
   Valinia, A. \& Marshall, F.~E. 1998, \apj\vol{505}{134}
\reference{}
   Webber, W.~R. 1997, \ssr\vol{81}{107}
\reference{}
   Weizs\"acker, C.~F.~V. 1934, Zeit. f\"ur Physik \vol{88}{612}
\reference{}
   Williams, E.~J. 1935, Kongelige Danske Vid. Selsk. Mat. Fys. Medd.
   \vol{13}{No. 4}
\reference{}
   Wright, E.~L., et al. 1994, \apj\vol{420}{450}
\reference{}
   Zimmermann, H.~U., Tr\"umper, J.~E., \& Yorke, H. 1996,
   R\"ontgenstrahlung from the Universe, MPE Report 263
   (Max-Planck-Institut f\"ur Extraterrestrische Physik, Garching).
%
\end{references}
\end{document}